\documentclass[
aps,
prx,
showpacs,
preprintnumbers,
twocolumn,
superscriptaddress,
]{revtex4-2}

\newcommand{\app}{\approx}

\newcommand{\sq}[1]{\sqrt{#1}}

\newcommand{\rt}{\rightarrow}

\newcommand{\beq}{\begin{equation}}
\newcommand{\eeq}{\end{equation}}
                  
\newcommand{\benum}{\begin{enumerate}}
\newcommand{\eenum}{\end{enumerate}}
                    
\newcommand{\bit}{\begin{itemize}}
\newcommand{\eit}{\end{itemize}}
\newcommand{\xhat}{\hat{\T{x}}}
\newcommand{\yhat}{\hat{\T{y}}}
\newcommand{\zhat}{\hat{\T{z}}}

\newcommand{\bea}{\begin{eqnarray}}
\newcommand{\eea}{\end{eqnarray}}

\newcommand{\bev}{\begin{verbatim}}
\newcommand{\eev}{\end{verbatim}}

\newcommand{\zt}{\times}

\newcommand{\T}[1]{\textbf{#1}}
\newcommand{\I}[1]{\textit{#1}}
\newcommand{\R}[1]{\textrm{#1}}

\newcommand{\zfl}[1]{\protect\label{fig:#1}}
\newcommand{\zfr}[1]{\figurename\,\ref{fig:#1}}

\newcommand{\ba}{\left\{ \begin{array}{lr}}
\newcommand{\ea}{\end{array}\right.}

\newcommand{\blist}[1]{
 \begin{list}{#1}
 \begin{align}
	 arrow
 \end{align}
 $\checkmark\star
  { \setlength{\itemsep}{3pt}
     \setlength{\parsep}{2pt}
     \setlength{\topsep}{3pt}
     \setlength{\partopsep}{0pt}
     \setlength{\leftmargin}{1em}
     \setlength{\labelwidth}{1em}
     \setlength{\labelsep}{0.5em} } }
\newcommand{\elist}{
  \end{list}  }

\DeclareMathSymbol{\vartheta}{\mathalpha}{letters}{"12}
\DeclareMathSymbol{\theta}{\mathalpha}{letters}{"23}
\DeclareMathSymbol{\phi}{\mathalpha}{letters}{"27}
\DeclareMathSymbol{\varphi}{\mathalpha}{letters}{"1E}

\newcommand{\bef}
{
\begin{figure}[htbp]
\centering
}

\newcommand{\eef}{\end{figure}}

\usepackage[T1]{fontenc}
\usepackage{newtxtext}
\usepackage{newtxmath}
\usepackage[normalem]{ulem}

\usepackage{enumitem}

\usepackage{makecell}

\usepackage{physics}
\usepackage{amsmath}
\usepackage{mathtools}
\usepackage{dsfont}

\usepackage{comment}

\usepackage{appendix}

\usepackage{placeins}

\usepackage{color}
\usepackage[dvipsnames]{xcolor}
\usepackage{colortbl}
\usepackage{hyperref}
\hypersetup{colorlinks,citecolor=blue,linkcolor=blue,urlcolor=blue}

\definecolor{darkgreen}{rgb}{0.55, 0.71, 0.00}
\definecolor{Gray}{gray}{0.9}

\usepackage{xpatch}

\makeatletter
\patchcmd{\@ssect@ltx}
    {\addcontentsline{toc}{#1}{\protect\numberline{}#8}}
    {}
    {}
    {}
\makeatother
\usepackage{xspace}
\usepackage{layouts}

\usepackage{graphicx}
\usepackage[all]{hypcap}

\usepackage{tikz} 

\graphicspath{{FiguresSI/}{FiguresMain/}}

\renewcommand{\figurename}{Fig.}

\newcommand{\affA}{Department of Chemistry, University of California, Berkeley, Berkeley, CA 94720, USA.}

\newcommand{\affE}{Chemical Sciences Division,  Lawrence Berkeley National Laboratory,  Berkeley, CA 94720, USA.}
\newcommand{\affF}{CIFAR Azrieli Global Scholars Program, 661 University Ave, Toronto, ON M5G 1M1, Canada.}
\newcommand{\affG}{Department of Pure and Applied Science, University of Urbino Carlo Bo, Urbino, I-61029, Italy.} 
\newcommand{\affH}{Molecular Foundry,  Lawrence Berkeley National Laboratory,  Berkeley, CA 94720, USA.}
\newcommand{\affI}{Department of Physics, Guru Nanak Dev University, Amritsar, Punjab 143102, India.}

\begin{document}
\title{High sensitivity pressure and temperature quantum sensing in organic crystals}

\author{Harpreet Singh}\affiliation{\affA}\affiliation{\affI}
\author{Noella D'Souza}\affiliation{\affA}\affiliation{\affE}
\author{Joseph Garrett}\affiliation{\affA}
\author{Angad Singh}\affiliation{\affA}
\author{Brian Blankenship}\affiliation{\affA}
\author{Emanuel Druga}\affiliation{\affA}
\author{Riccardo Montis}\affiliation{\affG}
\author{Liang Tan}\affiliation{\affH}
\author{Ashok Ajoy}\affiliation{\affA}\affiliation{\affE}\affiliation{\affF}

\begin{abstract}
The inherent sensitivity of quantum sensors to their physical environment can make them good reporters of parameters such as temperature, pressure, strain, and electric fields. Here, we present a molecular platform for pressure (P) and temperature (T) sensing using para-terphenyl crystals doped with pentacene. We leverage the optically detected magnetic resonance (ODMR) of the photoexcited triplet electron in the pentacene molecule, that serves as a sensitive probe for lattice changes in the host para-terphenyl due to pressure or temperature variations. We observe maximal ODMR frequency variations of $df/dP{=}1.8$ MHz/bar and $df/dT{=} 247$ kHz/K, which are over 1,200 times and three times greater, respectively, than those seen in nitrogen-vacancy centers in diamond. This results in a >85-fold improvement in pressure sensitivity over best previously reported. The larger variation reflects the weaker nature of the para-terphenyl lattice, with first-principles DFT calculations indicating that even picometer-level shifts in the molecular orbitals due to P, T changes are measurable. The platform offers additional advantages including high levels of sensor doping, narrow ODMR linewidths and high contrasts, and ease of deployment, leveraging the ability for large single crystals at low cost. Overall, this work paves the way for low-cost, optically-interrogated pressure and temperature sensors and lays the foundation for even more versatile sensors enabled by synthetic tunability in designer molecular systems. 
\end{abstract}

\maketitle
\pagebreak

\begin{figure}[t]
  \centering
  {\includegraphics[width=0.49\textwidth]{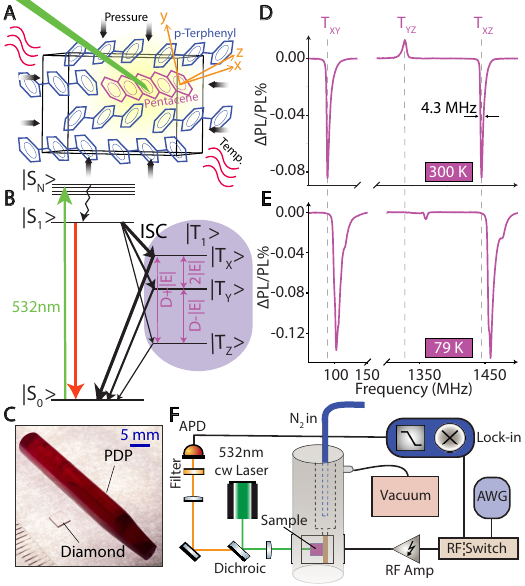}}
  \caption{\T{System and Principle.}
(A) \I{System}: Unit cell of optically-interrogated pentacene doped \I{p}-terphenyl (PDP). Molecular principal axes are denoted as $\xhat, \yhat, \zhat$. Red wavy lines and black arrows denote applied changes in temperature or pressure.
(B) \I{Energy level diagram} of the pentacene molecule, showing a singlet state $\ket{S_0}$ and a metastable triplet manifold $\ket{T_1}$ defined by ZFS parameters $D$ and $E$. Arrow thickness schematically illustrate the differential rates of $\ket{T_1}$ sub-level population and depopulation.
(C) \I{Low-cost cm-scale} PDP crystals are shown. A typical commercial NV-diamond sample is shown alongside for comparison (see scale bar).
(D-E) \I{Representative ODMR spectra} at zero-field and at (D) 300K and (E) 79K, showing marked triplet transitions with narrow spectral lines. Peaks blue shift as temperature decreases. Vertical dashed lines serve as a visual guide.
(F) \I{Experimental setup} includes a cryostat or pressure chamber housing the sample, a 532 nm CW laser for illumination, and fluorescence collection into an APD via a dichroic mirror. Microwaves are produced by an AWG and delivered via a self-shorted loop.}
\zfl{mfig1}
\end{figure}

\section{Introduction}
Quantum sensors are revolutionizing the precise measurement of various physical quantities because of their inherent sensitivity to their environment~\cite{Degen17}. While sensors constructed from electronic spins, such as Nitrogen Vacancy (NV) centers in diamond~\cite{Doherty12, Jelezko06}, are widely used as sensitive magnetic field sensors~\cite{Taylor08, Wolf15}, there is growing interest in their ability to probe other parameters, particularly temperature~\cite{Neumann13,Doherty14}, pressure~\cite{Cai14}, strain~\cite{Udvarhelyi18,Ovartchaiyapong14}, electric field~\cite{Dolde11}, and rotation~\cite{Ajoy2012stable,ledbetter2012gyroscopes}.

Sensing with NV centers and other quantum sensors leverages the sensitivity of the triplet-state zero-field splitting (ZFS), \( D \), to temperature, pressure or strain, enabling their local, sub-micron-scale, measurement~\cite{Balasubramanian08,Maletinsky12}. Applications include nanoscale thermometry in single cells~\cite{Kucsko13,Fujiwara20,Fujiwara21} and probing phase transitions of condensed matter systems in high-pressure anvil cells~\cite{Lesik19,Hsieh19,Hilberer23}. Optical interrogation of these sensors enables diffraction-limited, non-invasive sensing — capabilities often lacking in classical sensors (e.g. thermocouples).

Material properties, however, impose an overall bound on achievable sensitivity. For diamond NV centers the slope of variations with temperature and pressure are respectively, \( \frac{\partial D}{\partial T} {=} 71~ \)kHz/K and \( \frac{\partial D}{\partial P} {=} 1.46~ \)kHz/bar~\cite{Acosta10,Doherty14L}; the rigidity of the diamond lattice results in a relatively weak pressure (and strain) sensitivity~\cite{Ivady14,Udvarhelyi18}. Downstream implications include an increase in the technical complexity required for manipulating spins via strain~\cite{Macquarrie13,Ovartchaiyapong14,Meesala16,Bennett13}. This motivates exploration of alternative materials that also host a spin-optical interface, similar to NV centers, while offering an enhanced sensitivity to these physical parameters.

Recent advances have highlighted the potential of \I{molecular} systems for quantum sensing, utilizing are-earth ions~\cite{Bayliss20,Serrano22} or photoexcited organic radicals~\cite{Harvey21,Xie23}. These systems offer advantages stemming from bottom-up synthesis~\cite{Zadrozny17},  anticipating chemical control and tunable sensor placement in three-dimensions, and offering a pathway to customizing sensor properties at the molecular level.

As a prototypical unit, we recently demonstrated that pentacene molecules doped in para (\I{p})-terphenyl exhibit excellent spin-optical properties and can be exploited for optical magnetometry at room temperature (RT)~\cite{Mena24,Singh24}. The photoexcited triplet electron spin can be optically initialized and possesses state-dependent fluorescence contrast, yielding narrow-linewidth optically detected magnetic resonance (ODMR) spectra at RT~\cite{Singh24}. Additionally, the material can be grown into large single crystals with high doping levels, relative to defects in semiconductor materials, and low concentration of background paramagnetic impurities.

In comparison to defects in semiconductor materials like diamond, the weak, easily deformable, \I{p}-terphenyl lattice suggests that this material might exhibit heightened sensitivity to pressure and temperature. In this paper, we show this through a systematic study of photoexcited triplet ODMR spectra across a wide range of temperatures (77-330K) and pressures (0-8 bar). We measure a pressure and temperature slope ${>}$1200-fold and ${>}$3-fold greater than that of NV centers respectively, besides other operational advantages. First-principles DFT calculations support experimental findings, provide insight into origins of the enhanced sensitivity, and suggest potential for further improvements in designer chemical systems.

\section{System and Principle}
The sample is a single-crystal of pentacene doped \I{p}-terphenyl (PDP), doped at the 0.1\% level. \zfr{mfig1}A illustrates the lattice structure; $\xhat$ denotes the molecular long-axis, with $\yhat,\zhat$ transverse to it. Crystals were grown using the Bridgman technique~\cite{Bridgman64, Jiang13} after zone-refining the \I{p}-terphenyl host and subliming pentacene for purification (see SI). Doping levels exceed those of defects in semiconductor materials (e.g. diamond) by at least two orders of magnitude.  The PDP crystals can be grown up to several \I{cm} at low cost. \zfr{mfig1}C compares typical PDP crystal sizes with diamond. The PDP crystal in \zfr{mfig1}C requires only \$2.25 in materials cost, representing a ${\sim}$70,000-fold reduction in mass-normalized cost compared to NV-diamond. Polycrystalline material can be obtained by crushing single crystals or inducing imperfect growth to form mm-scale domains.

\zfr{mfig1}B shows the energy level diagram of the pentacene $\pi$-electron in the \I{p}-terphenyl host. It includes a ground state singlet ($\ket{S_0}$), an excited state singlet ($\ket{S_1}$), and a metastable triplet state ($\ket{T_1}$) represented by $\ket{T_x}, \ket{T_y}, \ket{T_z}$, with lifetimes of ${\sim}$35, 166, and 500 $\mu$s~\cite{wu2019unraveling}. The photoexcited triplet state is described by the spin Hamiltonian ${\cal H}_{\R{sys}}{=}D\left(S_{z}^{2}-\frac{2}{3}\right)+E(S_{x}^{2}-S_{y}^{2})$, where $\boldsymbol{S}$ is a spin-1 Pauli operator, with ZFS parameters $D{\app}1392$ MHz and $E{\app}53$ MHz~\cite{Yang00}.

Optical excitation populates the triplet state via intersystem crossing (ISC) as $\ket{S_1}{\rt}\ket{T_{x,y,z}}$ (see \zfr{mfig1}B); with pulsed excitation, the $\ket{T_x}$ state is polarized to ${\app}$76\%~\cite{Takeda02}. By selecting appropriate delays, spin state-dependent fluorescence contrast can be obtained exploiting differential relaxation from $\ket{T_1}{\rt}\ket{S_0}$, in a manner that is conditional on the triplet sub-levels~\cite{Singh24,Mena24}. This produces an ODMR spectrum (\zfr{mfig1}D-E). \zfr{mfig1}D the case for RT and Earth's field. Three transitions are marked, hosting narrow linewidths, here $\ell_0{\app}4.3$ MHz, even at the high doping level and with power broadening~\cite{Singh24}. The spectra here are with illumination with a CW laser; pulsed laser excitation can yield higher ODMR contrast (${\sim}$17\%)~\cite{singh2024room}. Asymmetric lineshapes are influenced by hyperfine couplings to neighboring proton nuclei. Inversion of contrast for the $T_{yz}$ $(\vert T_y\rangle {\leftrightarrow} \vert T_z\rangle$) transitions is due to higher steady-state population in $\ket{T_z}$ than $\ket{T_y}$ (see SI). For $T_{xy}$, the electronic coherence time was measured at $T_2^{\R{DD}}{\app}18\mu$s under dynamic decoupling, with $T_1{=}23\mu$s, dominated by triplet-ground relaxation~\cite{Singh24}. 

As schematically shown in \zfr{mfig1}A, we investigate changes in the ODMR spectra under varying temperature or pressure, which affect the host lattice and the ZFS parameters, $D$ and $E$. \zfr{mfig1}F illustrates the experimental setup. The sample is placed in a variable-temperature flow cryostat (Janis ST100) or a pressure chamber. Experiments are conducted on a sub-ensemble of ${\sim}10^9$ pentacene molecules over $2.6{\zt}10^{-5}$ mm$^3$.

\zfr{mfig1}E shows a typical result. Compared to the RT ODMR spectrum (\zfr{mfig1}D), lowering the temperature to 77K causes a noticeable shift in the positions of the three transitions, as highlighted by the dashed vertical lines.

\begin{figure*}[t]
  \centering
  {\includegraphics[width=0.97\textwidth]{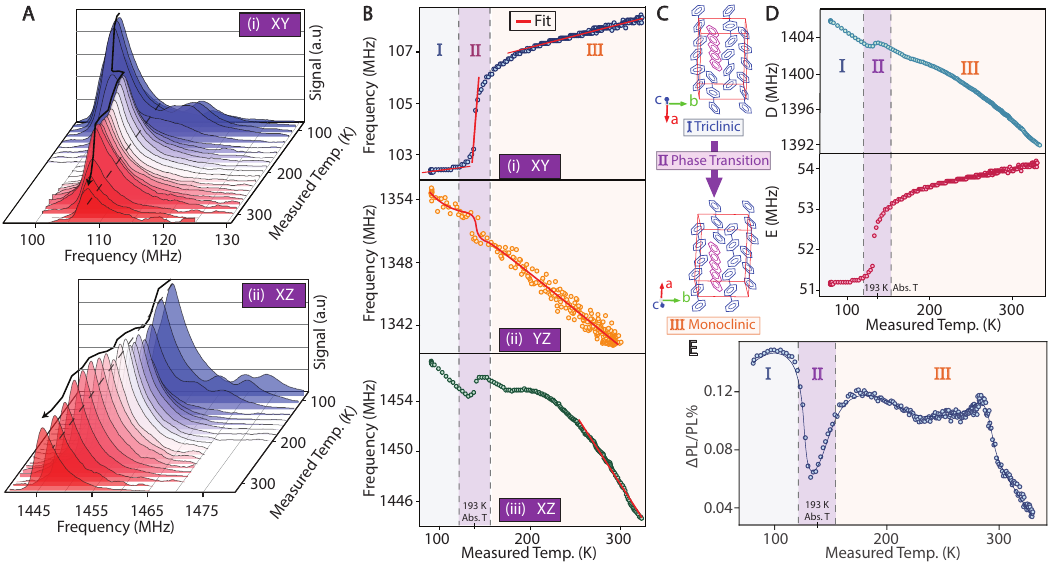}}
  \caption{\T{Triplet ODMR Temperature Sensing.} 
  (A) \I{ODMR spectra} of $T_{xy}$ and $T_{xz}$ transitions with changing temperature. Reported are values from the cryostat cold-finger, and do not accounting for sample heating. Blue-to-red colors represent increasing temperature. Dashed line is guide to eye; black arrow tracks changes in spectral peak positions. 
(B) \I{Temperature variation} of ODMR peak position for the (i) $T_{xy}$, (ii) $T_{yz}$ and (i) $T_{xz}$ transition over a wide range (77 to 330 K). Three linear regions are observed, marked \T{I}-\T{III}, with distinct slopes. Sharp variation in region \T{II} is due to a phase transition, at an absolute temperature of 193K~\cite{baer93temp} (marked).
(C) \I{Lattice phases} corresponding to regions \T{I} and \T{III} are identified as triclinic and monoclinic, respectively. 
(D) \I{Temperature dependence} of the zero-field splitting parameters, $D(T)$ and $E(T)$.
(E) \I{ODMR contrast} extracted over the temperature range for the $T_{xy}$ transition. General contrast increase is observed at lower temperatures, with sharp contrast variation near the phase transition, and a decrease for $T{>}290$K.
}
\zfl{mfig2}
\end{figure*}

\begin{table}
    \centering
    \includegraphics[width=1\linewidth]{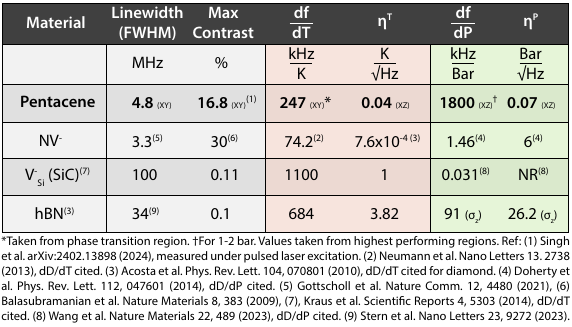}
    \caption{\T{Comparison of quantum sensor platforms} for temperature and pressure sensing. Source references are shown as footnotes. Slopes of ODMR frequency variations $df/dT$ and $df/dP$ which are material properties; while sensitivity values $\eta^T,\eta^P$ depend on particulars of the measurement. For pentacene, subscripts identify triplet transition from where values are extracted, and sensitivity values are under currently demonstrated conditions. First two columns shown ODMR linewidth and contrast but are not employed for sensitivity estimation.}
\label{table:sensitivity} 
\end{table}

\begin{figure}
  \centering
  
  {\includegraphics[width=0.49\textwidth]{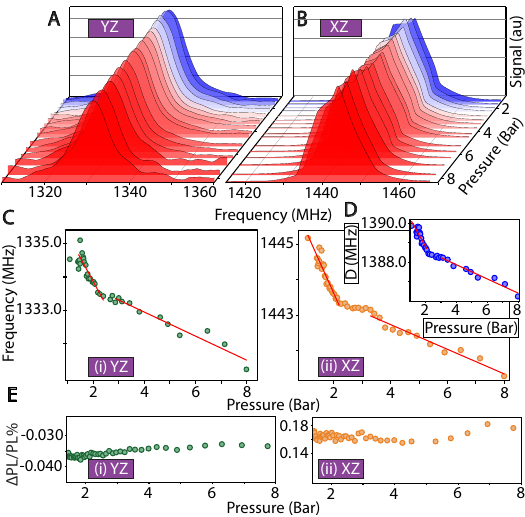}}
  \caption{\T{Triplet ODMR Isotropic Pressure Sensing}
  (A) \I{Representative ODMR traces} for the (A) $T_{yz}$ and (B) $T_{yz}$ transitions under varying isotropically applied pressure 0-8bar. Colors blue-to-red indicate increasing pressure.  (C) \I{Pressure variation} of the ODMR peak frequency for the (i) $T_{yz}$ and (ii) $T_{xz}$ transitions, both exhibit high sensitivity (see Table~\ref{table:sensitivity}). Red lines indicate characteristic linear fits.
(D) \I{Pressure dependence of ZFS parameter} $D(P)$ extracted from (C). 
(E) \I{Variation in ODMR contrast} for (i) $T_{yz}$ and (ii) $T_{xz}$ transitions. Contrast remains approximately constant over the range studied. 
}
\zfl{mfig3}
\end{figure}

\section{Triplet ODMR variation with temperature}
To investigate the spectral changes with temperature, \zfr{mfig2} presents data across a wide range (77K-330K). \zfr{mfig2}A(i-ii) shows individual ODMR traces for $T_{xy}$ and $T_{xz}$, with color gradients (blue-to-red) representing increasing temperatures. The temperatures are measured at the cryostat cold-finger and have a constant offset from the actual sample temperature due to laser induced heating. A dashed line parallel to the temperature axis serves as a visual guide, making the spectral shift immediately apparent. The movement of the peaks is indicated by the black arrows.

\zfr{mfig2}B(i) shows the extracted ODMR peak positions for $T_{xy}$ transition estimated from the center of the steep spectral edge, plotted against the measured cold-finger temperature. The data reveal three distinct linear regions, labeled $\T{I}-\T{III}$ and shaded for clarity. The strong variation around region \T{II}) has the characteristic signature of a phase transition in the \I{p}-terphenyl molecules~\cite{Rice13}. From the literature, this phase transition occurs at 193 K~\cite{baer93temp}; this is marked in \zfr{mfig2}B for clarity. 

\zfr{mfig2}C schematically depicts the lattice diagrams of the two phases, transitioning from triclinic in region \T{I} to monoclinic in region \T{III}. The structures are similar, except for the central \I{p}-terphenyl benzene ring, which is out-of-plane in the triclinic phase. While related signatures have been observed previously in photoluminescence~\cite{Croci93}, to our knowledge, \zfr{mfig2}B marks the first time an ODMR measurement is carried out directly at a phase transition.  Overall, \zfr{mfig2}B demonstrates that pentacene molecules are sensitive spectators to changes in the host lattice configuration. 

The red lines in \zfr{mfig2}B(i) show linear fits to the ODMR variation in the three region. The slope in region \T{II} is about three times that of the variation in diamond (see Table \ref{table:sensitivity} for a detailed comparision). As this phase transition is reversible, it may serve as an excellent bias point for quantum sensing thermometry. We anticipate another sharp, albeit irreversible, phase transition at the melting point around 486K.

\zfr{mfig2}B(ii-iii) shows corresponding variations for the $T_{yz}$ and $T_{xz}$ transitions respectively. The step variation at the phase transition is visible in both cases. $T_{yz}$ exhibits an approximately linear dependence over the entire temperature range studied, and constitutes a wider linear dynamic range than in NV-diamond~\cite{Jarmola12}. 
From a practical perspective, the complementary use of the steep $T_{xy}$ transition and the linear $T_{yz}$ transition allows for both high sensitivity and a large dynamic range temperature sensing within the same system.

\zfr{mfig2}D presents the extracted changes in the ZFS parameters with temperature, $D(T)$ and $E(T)$. Table \ref{table:sensitivity} compares these variations with other quantum sensing materials, including NV-diamond, silicon vacancies (V\(_\text{Si}^-\)) in silicon carbide (SiC) (shown are values for the excited state), and negatively charged boron vacancies (V\(_\text{B}^-\)) in hBN. The third column, $df/dT$, represents the change in spectral frequency with temperature, a material-specific parameter independent of measurement apparatus or light collection efficiency. For pentacene, we observe a variation of 247kHz/K for $T_{xy}$ in region \T{II} and 101kHz/K for $T_{xz}$ in region \T{III} (red line in \zfr{mfig2}B(iii)), both steeper than the variation in NV-diamond.

\zfr{mfig2}E now examines variations in ODMR contrast, focusing on the $T_{xy}$ transition (see SI for other transitions). Contrast increases slightly at lower temperatures but shows a sharp change near the phase transition in region \T{II}. Another drop occurs after the plateau past 290K, likely due to the crystal getting closer to its melting point. Notably, much higher absolute contrast (up to 17\%) can be achieved using pulsed laser excitation~\cite{Singh24}, and we expect similar contrast variations as shown in \zfr{mfig2}E even in this case.

To evaluate the time-normalized temperature sensitivity of our measurements, we use $\eta^{T} {=} \sigma \sqrt{\tau} / \frac{dS}{dT}$\cite{CHOE20181066}, where $\frac{dS}{dT}$ is the maximum ODMR signal slope with temperature, $\sigma$ is the noise floor, and $\tau$ is the integration time, defined by the low-pass filter’s settling time in the detection lock-in amplifier. Our setup is not optimized for sensitivity; we collect only a small fraction of photons, and the ODMR contrast in \zfr{mfig2}E is low due to continuous-wave illumination. Both factors could be improved by at least an order of magnitude~\cite{singh2024room}. 

Even so, in region \T{II} of the $T_{xz}$ transition (\zfr{mfig2}B(iii)), we estimate a sensitivity of $\eta^{T}{=}0.04$kHz/K. The fourth column of Table \ref{table:sensitivity} shows the best reported values from other systems. A direct comparison is challenging, as sensitivity depends on many measurement parameters, such as photon collection and the number of spins interrogated. Nevertheless, even with our current setup, pentacene outperforms defects in SiC and hBN, partially due to the narrower ODMR linewidth~\cite{kraus2014magnetic} (first column in Table \ref{table:sensitivity}). With straightforward improvements, we anticipate approaching NV-diamond sensitivity; already however pentacene already offers significant deployment advantages due to the ease of crystal growth and lower cost (see \zfr{mfig1}C).

\section{Triplet ODMR variation with pressure}
An analogous study was conducted for ODMR variation with applied isotropic pressure, shown in \zfr{mfig3} for a low absolute pressure range (0-8 bar). \zfr{mfig3}A-B presents representative ODMR traces for the $T_{yz}$ and $T_{xz}$ transitions. The traces show minimal spectral broadening with applied pressure, along with a discernible shift in the peak position. In contrast to NV-diamond which is more suited to high-bias pressure environments, \zfr{mfig3}A-B demonstrates the ability for measurements at close to ambient conditions.

\zfr{mfig3}C shows the variation in ODMR transition frequencies for the (i) $T_{yz}$ and (ii) $T_{xz}$ transitions, similar to \zfr{mfig2}B. As expected, no lattice phase transition occurs within the pressure range studied. The variation in \zfr{mfig3}C is weakly nonlinear, but for simplicity, we estimate two linear slopes over the measured range, indicated by the red lines. For the $T_{xz}$ transition, we estimate a  $df/dP$ variation of 1.8 MHz/Bar in the (1-2 Bar) range and 350 kHz/Bar in the (3-8 Bar) range. Variation for the $T_{yz}$ transition is similar: 1.4 MHz/Bar in the (1-2 Bar) range and 362 kHz/Bar in the (3-8 Bar) range.

As shown in the fifth column of Table \ref{table:sensitivity}, the maximum variation here is at least 1200 times greater than that of NV centers in diamond, and even larger for the case of V\(_\text{Si}^-\) in SiC. This difference can be attributed to the relative weakness of the \I{p}-terphenyl lattice; also reflected in the lower Young’s modulus in \I{p}-terphenyl (70 kBar~\cite{Heimel03}) compared to diamond (12 MBar) and SiC (4.5 MBar)~\cite{Baimova04}. \zfr{mfig3}D presents the extracted $D(P)$ parameter, while $E(P)$ remains approximately constant over the range studied (see SI)~\cite{Baer94}.

\zfr{mfig3}E shows the variation in ODMR contrast over the pressure range, for the (i) $T_{yz}$ and (ii) $T_{xz}$ transitions respectively. The contrast is approximately constant throughout the range. Under the current conditions, time-normalized pressure sensitivity can be evaluated as $\eta^{P} {=} \sigma \sqrt{\tau} / \frac{dS}{dP}$, and is reported in the sixth column in Table~\ref{table:sensitivity}. Even without optimization, the pressure sensitivity for PDP (${\app}$0.07Bar/$\sq{\R{Hz}}$) significantly outperforms the best reported values for other platforms, while operating in a convenient range near ambient pressure.

\begin{figure}
  \centering
  {\includegraphics[width=0.49\textwidth]{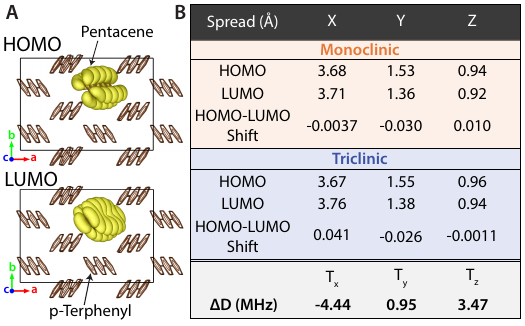}}
  \caption{\T{Pentacene HOMO/LUMO Character from DFT} 
 (A) 3D renderings of the HOMO and LUMO molecular orbitals, indicated in yellow, in pentacene. (B) Pentacene HOMO and LUMO parameters in monoclinic and triclinic \I{p}-terphenyl phases, obtained from DFT calculations. HOMO/LUMO spread is the standard deviation of the pentacene HOMO/LUMO orbital density $\vert\phi\vert^2$. HOMO-LUMO shift is the difference of the centroid of the orbital densities. $\Delta\mathbf{D}$ is the change in zero field splitting tensor eigenvalues between the monoclinic and triclinic phases, for the $T_1$ exciton in the $\xhat$, $\yhat$, $\zhat$ directions.
}  
\zfl{mfig4}
\end{figure}

\section{DFT calculations}
To understand these trends, we perform density functional theory calculations using a plane-wave basis set and norm-conserving pseudopotentials as implemented in the Quantum ESPRESSO code~\cite{giannozzi_quantum_2009}. We use a kinetic energy cutoff of 60 Ry, and adopt spin collinear Perdew-Burke-Ernzerhof (PBE) as the exchange-correlation energy functional. Initial crystal structures of monoclinic and triclinic p-terphenyl were obtained from Refs.~\cite{rice_temperature_2013,rietveld_x-ray_1970}. One pentacene molecule is substituted into the M4 position~\cite{guttler_single_1996}, and the ground state structures are further optimized until the Hellmann-Feynman forces on each atom are smaller than $10^{-5}$ au in magnitude. We use a 1${\times1\times} 2$ supercell containing 1 pentacene molecule and 15 p-terphenyl molecules. Brillouin-zone integrations are sampled on a uniform grid of 2$\times2\times 1$ k-points. To simulate the optically excited triplet state, we follow the $\Delta$SCF method~\cite{hellman_potential-energy_2004}, fixing the occupations so that the highest occupied molecular orbital (HOMO) of pentacene contains one spin-up electron and the lowest unoccupied molecular orbital (LUMO) of pentacene contains one spin-up electron. Renderings of these orbitals are shown in \zfr{mfig4}A.

We focus here on the restricted question of computing the change $\Delta D$ corresponding to the phase change between monoclinic and triclinic in \zfr{mfig2}B and \zfr{mfig2}D. Between the monoclinic and triclinic phases, we find relative shifts between the pentacene HOMO and LUMO centroid position as well as decreases in the spread of these orbitals. Orbital spreads are generally smaller in the monoclinic phase by $1{\sim}5$pm (\zfr{mfig4}B). Lattice constants in the monoclinic phase are larger than the triclinic phase by about 0.2\AA\ in the $a, c$ directions, leading to reduced intermolecular interactions and tighter localization of molecular orbitals. Because of the crystal field, pentacene orbitals are not perfectly centered on the molecule. In particular, orbital centroids shift by $1{\sim}4$pm  between monoclinic and triclinic phases, with the largest relative shift along the long molecular axis ($\xhat$).  

These changes in the molecular orbitals lead to differences in the zero-field splitting tensor
 between the two phases. The spin-spin interaction Hamiltonian \cite{harriman1978} is 
\begin{equation}
    \mathbf{D}_{ab} = \frac{1}{2}\frac{\mu_0}{4\pi}(g_e\mu_B)^2\sum_{i<j}\chi_{ij} \langle \Phi_{ij} \vert \frac{r^2\delta_{ab}-3r_ar_b}{r^5}\vert \Phi_{ij}\rangle
\end{equation}
for all electron pairs $\Phi_{ij}(r,r') {=} \frac{1}{\sqrt{2}}(\phi_i(r)\phi_j(r')-\phi_j(r)\phi_i(r'))$ and $\chi_{ij}{=}\pm1$ for parallel($+$)/anti-parallel($-$) electrons.  Considering just the contribution from the pentacene HOMO and LUMO orbitals, we find differences in $\mathbf{D}_{ab}$ eigenvalues between the monoclinic and triclinic phases of up to 4 MHz (\zfr{mfig4}B), which is the same order of magnitude as the experimentally observed frequency shifts (\zfr{mfig2}B). 

Furthermore, the $T_x$ frequency shifts the most, as a result of the larger orbital shifts along the $\xhat$ direction, agreeing with the experimental observation of $T_{xy}$ and $T_{xz}$ transitions shifting the most.  Including contributions of all 1392 electrons in the system to $\mathbf{D}_{ab}$ is computationally unfeasible. We expect these contributions to change the exact values of frequency shifts, while remaining at the same order of magnitude. We may estimate the change in the spin-spin interaction as $\Delta\mathbf{D}\approx \frac{1}{2}\frac{\mu_0}{4\pi}(g_e\mu_B)^2 \frac{\Delta r}{r^4}$. Taking typical values from \zfr{mfig4}B of the change in localization length $\Delta r{= }4$ pm, and the orbital spread $r{\approx}3.7$\AA, we estimate $\Delta\mathbf{D}{\approx} 1$ MHz. This analysis shows that picometer scale changes in molecular orbitals are measurable by ODMR peak shifts at the MHz scale for such systems.    

\section{Discussion and Outlook}
Our work suggests several intriguing future directions. As Table~\ref{table:sensitivity} highlights, pentacene-doped \I{p}-terphenyl crystals are  compelling for pressure and temperature sensing. Other advantages, such as the ability to grow large (\I{cm}-scale) crystals (\zfr{mfig1}C) and easily cleave them, suggest the potential for large-area $P$, $T$ sensor arrays. The crystals are free of paramagnetic impurities (unlike diamond P1 centers~\cite{Reynhardt98}) allowing intrinsically large sensor densities. Table~\ref{table:sensitivity} indicates that these materials are particularly suitable for high-temperature dynamic range and low-bias pressure environments, a complementary regime to diamond and SiC, which are better suited to high-bias pressure settings.

More broadly, our work highlights the benefits of chemical systems for $P$, $T$ quantum sensing. This approach eliminates the reliance on electronic defects in semiconductor lattices and opens new design possibilities through chemical synthesis~\cite{Bayliss20}. We anticipate increasing sensitivity by incorporating these molecules into porous materials such as metal-organic frameworks (MOFs)~\cite{Zadrozny17}, where higher structural flexibility can result in greater sensitivity to pressure and strain. This also suggests new quantum sensor form factors, including thin films~\cite{Mena24}, 3D printed materials~\cite{Blankenship23,Blankenship24}, and nanoparticles, possibly down to the single-molecule level~\cite{ambrose1991detection}. This anticipates single cell-deployable molecular temperature and strain sensing tags.

The large spin-strain coupling in these materials and their ease of fabrication significantly reduces the technical barrier to mechanically actuating the electronic spins~\cite{Macquarrie13}, for instance via micromechanical structures. It also presents a novel pathway to linearly shift resonance frequencies of individual molecules via strain~\cite{Stannigel10,Guo23,Lee16}, suggesting a method to individually address qubits in molecular quantum computing and sensing platforms~\cite{Wasielewski20}.

We gratefully acknowledge discussions with J. Breeze and S. Bhave, and funding from NSF TAQS,  ONR (N00014-20-1-2806), AFOSR YIP (FA9550-23-1-
0106), the Noyce Foundation, and the CIFAR Azrieli Foundation (GS23-013). 

\vspace{-5mm} 

\vspace{-1mm}


\begin{thebibliography}{63}%
  \makeatletter
  \providecommand \@ifxundefined [1]{%
   \@ifx{#1\undefined}
  }%
  \providecommand \@ifnum [1]{%
   \ifnum #1\expandafter \@firstoftwo
   \else \expandafter \@secondoftwo
   \fi
  }%
  \providecommand \@ifx [1]{%
   \ifx #1\expandafter \@firstoftwo
   \else \expandafter \@secondoftwo
   \fi
  }%
  \providecommand \natexlab [1]{#1}%
  \providecommand \enquote  [1]{``#1''}%
  \providecommand \bibnamefont  [1]{#1}%
  \providecommand \bibfnamefont [1]{#1}%
  \providecommand \citenamefont [1]{#1}%
  \providecommand \href@noop [0]{\@secondoftwo}%
  \providecommand \href [0]{\begingroup \@sanitize@url \@href}%
  \providecommand \@href[1]{\@@startlink{#1}\@@href}%
  \providecommand \@@href[1]{\endgroup#1\@@endlink}%
  \providecommand \@sanitize@url [0]{\catcode `\\12\catcode `\$12\catcode `\&12\catcode `\#12\catcode `\^12\catcode `\_12\catcode `\%12\relax}%
  \providecommand \@@startlink[1]{}%
  \providecommand \@@endlink[0]{}%
  \providecommand \url  [0]{\begingroup\@sanitize@url \@url }%
  \providecommand \@url [1]{\endgroup\@href {#1}{\urlprefix }}%
  \providecommand \urlprefix  [0]{URL }%
  \providecommand \Eprint [0]{\href }%
  \providecommand \doibase [0]{https://doi.org/}%
  \providecommand \selectlanguage [0]{\@gobble}%
  \providecommand \bibinfo  [0]{\@secondoftwo}%
  \providecommand \bibfield  [0]{\@secondoftwo}%
  \providecommand \translation [1]{[#1]}%
  \providecommand \BibitemOpen [0]{}%
  \providecommand \bibitemStop [0]{}%
  \providecommand \bibitemNoStop [0]{.\EOS\space}%
  \providecommand \EOS [0]{\spacefactor3000\relax}%
  \providecommand \BibitemShut  [1]{\csname bibitem#1\endcsname}%
  \let\auto@bib@innerbib\@empty
  \bibitem [{\citenamefont {Degen}\ \emph {et~al.}(2017)\citenamefont {Degen}, \citenamefont {Reinhard},\ and\ \citenamefont {Cappellaro}}]{Degen17}%
    \BibitemOpen
    \bibfield  {author} {\bibinfo {author} {\bibfnamefont {C.~L.}\ \bibnamefont {Degen}}, \bibinfo {author} {\bibfnamefont {F.}~\bibnamefont {Reinhard}},\ and\ \bibinfo {author} {\bibfnamefont {P.}~\bibnamefont {Cappellaro}},\ }\href@noop {} {\bibfield  {journal} {\bibinfo  {journal} {Reviews of modern physics}\ }\textbf {\bibinfo {volume} {89}},\ \bibinfo {pages} {035002} (\bibinfo {year} {2017})}\BibitemShut {NoStop}%
  \bibitem [{\citenamefont {Doherty}\ \emph {et~al.}(2012)\citenamefont {Doherty}, \citenamefont {Dolde}, \citenamefont {Fedder}, \citenamefont {Jelezko}, \citenamefont {Wrachtrup}, \citenamefont {Manson},\ and\ \citenamefont {Hollenberg}}]{Doherty12}%
    \BibitemOpen
    \bibfield  {author} {\bibinfo {author} {\bibfnamefont {M.}~\bibnamefont {Doherty}}, \bibinfo {author} {\bibfnamefont {F.}~\bibnamefont {Dolde}}, \bibinfo {author} {\bibfnamefont {H.}~\bibnamefont {Fedder}}, \bibinfo {author} {\bibfnamefont {F.}~\bibnamefont {Jelezko}}, \bibinfo {author} {\bibfnamefont {J.}~\bibnamefont {Wrachtrup}}, \bibinfo {author} {\bibfnamefont {N.}~\bibnamefont {Manson}},\ and\ \bibinfo {author} {\bibfnamefont {L.}~\bibnamefont {Hollenberg}},\ }\href@noop {} {\bibfield  {journal} {\bibinfo  {journal} {Physical Review B—Condensed Matter and Materials Physics}\ }\textbf {\bibinfo {volume} {85}},\ \bibinfo {pages} {205203} (\bibinfo {year} {2012})}\BibitemShut {NoStop}%
  \bibitem [{\citenamefont {Jelezko}\ and\ \citenamefont {Wrachtrup}(2006)}]{Jelezko06}%
    \BibitemOpen
    \bibfield  {author} {\bibinfo {author} {\bibfnamefont {F.}~\bibnamefont {Jelezko}}\ and\ \bibinfo {author} {\bibfnamefont {J.}~\bibnamefont {Wrachtrup}},\ }\href@noop {} {\bibfield  {journal} {\bibinfo  {journal} {physica status solidi (a)}\ }\textbf {\bibinfo {volume} {203}},\ \bibinfo {pages} {3207} (\bibinfo {year} {2006})}\BibitemShut {NoStop}%
  \bibitem [{\citenamefont {Taylor}\ \emph {et~al.}(2008)\citenamefont {Taylor}, \citenamefont {Cappellaro}, \citenamefont {Childress}, \citenamefont {Jiang}, \citenamefont {Budker}, \citenamefont {Hemmer}, \citenamefont {Yacoby}, \citenamefont {Walsworth},\ and\ \citenamefont {Lukin}}]{Taylor08}%
    \BibitemOpen
    \bibfield  {author} {\bibinfo {author} {\bibfnamefont {J.~M.}\ \bibnamefont {Taylor}}, \bibinfo {author} {\bibfnamefont {P.}~\bibnamefont {Cappellaro}}, \bibinfo {author} {\bibfnamefont {L.}~\bibnamefont {Childress}}, \bibinfo {author} {\bibfnamefont {L.}~\bibnamefont {Jiang}}, \bibinfo {author} {\bibfnamefont {D.}~\bibnamefont {Budker}}, \bibinfo {author} {\bibfnamefont {P.}~\bibnamefont {Hemmer}}, \bibinfo {author} {\bibfnamefont {A.}~\bibnamefont {Yacoby}}, \bibinfo {author} {\bibfnamefont {R.}~\bibnamefont {Walsworth}},\ and\ \bibinfo {author} {\bibfnamefont {M.}~\bibnamefont {Lukin}},\ }\href@noop {} {\bibfield  {journal} {\bibinfo  {journal} {Nature Physics}\ }\textbf {\bibinfo {volume} {4}},\ \bibinfo {pages} {810} (\bibinfo {year} {2008})}\BibitemShut {NoStop}%
  \bibitem [{\citenamefont {Wolf}\ \emph {et~al.}(2015)\citenamefont {Wolf}, \citenamefont {Neumann}, \citenamefont {Nakamura}, \citenamefont {Sumiya}, \citenamefont {Ohshima}, \citenamefont {Isoya},\ and\ \citenamefont {Wrachtrup}}]{Wolf15}%
    \BibitemOpen
    \bibfield  {author} {\bibinfo {author} {\bibfnamefont {T.}~\bibnamefont {Wolf}}, \bibinfo {author} {\bibfnamefont {P.}~\bibnamefont {Neumann}}, \bibinfo {author} {\bibfnamefont {K.}~\bibnamefont {Nakamura}}, \bibinfo {author} {\bibfnamefont {H.}~\bibnamefont {Sumiya}}, \bibinfo {author} {\bibfnamefont {T.}~\bibnamefont {Ohshima}}, \bibinfo {author} {\bibfnamefont {J.}~\bibnamefont {Isoya}},\ and\ \bibinfo {author} {\bibfnamefont {J.}~\bibnamefont {Wrachtrup}},\ }\href@noop {} {\bibfield  {journal} {\bibinfo  {journal} {Physical Review X}\ }\textbf {\bibinfo {volume} {5}},\ \bibinfo {pages} {041001} (\bibinfo {year} {2015})}\BibitemShut {NoStop}%
  \bibitem [{\citenamefont {Neumann}\ \emph {et~al.}(2013)\citenamefont {Neumann}, \citenamefont {Jakobi}, \citenamefont {Dolde}, \citenamefont {Burk}, \citenamefont {Reuter}, \citenamefont {Waldherr}, \citenamefont {Honert}, \citenamefont {Wolf}, \citenamefont {Brunner}, \citenamefont {Shim} \emph {et~al.}}]{Neumann13}%
    \BibitemOpen
    \bibfield  {author} {\bibinfo {author} {\bibfnamefont {P.}~\bibnamefont {Neumann}}, \bibinfo {author} {\bibfnamefont {I.}~\bibnamefont {Jakobi}}, \bibinfo {author} {\bibfnamefont {F.}~\bibnamefont {Dolde}}, \bibinfo {author} {\bibfnamefont {C.}~\bibnamefont {Burk}}, \bibinfo {author} {\bibfnamefont {R.}~\bibnamefont {Reuter}}, \bibinfo {author} {\bibfnamefont {G.}~\bibnamefont {Waldherr}}, \bibinfo {author} {\bibfnamefont {J.}~\bibnamefont {Honert}}, \bibinfo {author} {\bibfnamefont {T.}~\bibnamefont {Wolf}}, \bibinfo {author} {\bibfnamefont {A.}~\bibnamefont {Brunner}}, \bibinfo {author} {\bibfnamefont {J.~H.}\ \bibnamefont {Shim}}, \emph {et~al.},\ }\href@noop {} {\bibfield  {journal} {\bibinfo  {journal} {Nano letters}\ }\textbf {\bibinfo {volume} {13}},\ \bibinfo {pages} {2738} (\bibinfo {year} {2013})}\BibitemShut {NoStop}%
  \bibitem [{\citenamefont {Doherty}\ \emph {et~al.}(2014{\natexlab{a}})\citenamefont {Doherty}, \citenamefont {Acosta}, \citenamefont {Jarmola}, \citenamefont {Barson}, \citenamefont {Manson}, \citenamefont {Budker},\ and\ \citenamefont {Hollenberg}}]{Doherty14}%
    \BibitemOpen
    \bibfield  {author} {\bibinfo {author} {\bibfnamefont {M.~W.}\ \bibnamefont {Doherty}}, \bibinfo {author} {\bibfnamefont {V.~M.}\ \bibnamefont {Acosta}}, \bibinfo {author} {\bibfnamefont {A.}~\bibnamefont {Jarmola}}, \bibinfo {author} {\bibfnamefont {M.~S.}\ \bibnamefont {Barson}}, \bibinfo {author} {\bibfnamefont {N.~B.}\ \bibnamefont {Manson}}, \bibinfo {author} {\bibfnamefont {D.}~\bibnamefont {Budker}},\ and\ \bibinfo {author} {\bibfnamefont {L.~C.}\ \bibnamefont {Hollenberg}},\ }\href@noop {} {\bibfield  {journal} {\bibinfo  {journal} {Physical Review B}\ }\textbf {\bibinfo {volume} {90}},\ \bibinfo {pages} {041201} (\bibinfo {year} {2014}{\natexlab{a}})}\BibitemShut {NoStop}%
  \bibitem [{\citenamefont {Cai}\ \emph {et~al.}(2014)\citenamefont {Cai}, \citenamefont {Jelezko},\ and\ \citenamefont {Plenio}}]{Cai14}%
    \BibitemOpen
    \bibfield  {author} {\bibinfo {author} {\bibfnamefont {J.}~\bibnamefont {Cai}}, \bibinfo {author} {\bibfnamefont {F.}~\bibnamefont {Jelezko}},\ and\ \bibinfo {author} {\bibfnamefont {M.~B.}\ \bibnamefont {Plenio}},\ }\href@noop {} {\bibfield  {journal} {\bibinfo  {journal} {Nature communications}\ }\textbf {\bibinfo {volume} {5}},\ \bibinfo {pages} {4065} (\bibinfo {year} {2014})}\BibitemShut {NoStop}%
  \bibitem [{\citenamefont {Udvarhelyi}\ \emph {et~al.}(2018)\citenamefont {Udvarhelyi}, \citenamefont {Shkolnikov}, \citenamefont {Gali}, \citenamefont {Burkard},\ and\ \citenamefont {P{\'a}lyi}}]{Udvarhelyi18}%
    \BibitemOpen
    \bibfield  {author} {\bibinfo {author} {\bibfnamefont {P.}~\bibnamefont {Udvarhelyi}}, \bibinfo {author} {\bibfnamefont {V.~O.}\ \bibnamefont {Shkolnikov}}, \bibinfo {author} {\bibfnamefont {A.}~\bibnamefont {Gali}}, \bibinfo {author} {\bibfnamefont {G.}~\bibnamefont {Burkard}},\ and\ \bibinfo {author} {\bibfnamefont {A.}~\bibnamefont {P{\'a}lyi}},\ }\href@noop {} {\bibfield  {journal} {\bibinfo  {journal} {Physical Review B}\ }\textbf {\bibinfo {volume} {98}},\ \bibinfo {pages} {075201} (\bibinfo {year} {2018})}\BibitemShut {NoStop}%
  \bibitem [{\citenamefont {Ovartchaiyapong}\ \emph {et~al.}(2014)\citenamefont {Ovartchaiyapong}, \citenamefont {Lee}, \citenamefont {Myers},\ and\ \citenamefont {Jayich}}]{Ovartchaiyapong14}%
    \BibitemOpen
    \bibfield  {author} {\bibinfo {author} {\bibfnamefont {P.}~\bibnamefont {Ovartchaiyapong}}, \bibinfo {author} {\bibfnamefont {K.~W.}\ \bibnamefont {Lee}}, \bibinfo {author} {\bibfnamefont {B.~A.}\ \bibnamefont {Myers}},\ and\ \bibinfo {author} {\bibfnamefont {A.~C.~B.}\ \bibnamefont {Jayich}},\ }\href@noop {} {\bibfield  {journal} {\bibinfo  {journal} {Nature communications}\ }\textbf {\bibinfo {volume} {5}},\ \bibinfo {pages} {4429} (\bibinfo {year} {2014})}\BibitemShut {NoStop}%
  \bibitem [{\citenamefont {Dolde}\ \emph {et~al.}(2011)\citenamefont {Dolde}, \citenamefont {Fedder}, \citenamefont {Doherty}, \citenamefont {N{\"o}bauer}, \citenamefont {Rempp}, \citenamefont {Balasubramanian}, \citenamefont {Wolf}, \citenamefont {Reinhard}, \citenamefont {Hollenberg}, \citenamefont {Jelezko} \emph {et~al.}}]{Dolde11}%
    \BibitemOpen
    \bibfield  {author} {\bibinfo {author} {\bibfnamefont {F.}~\bibnamefont {Dolde}}, \bibinfo {author} {\bibfnamefont {H.}~\bibnamefont {Fedder}}, \bibinfo {author} {\bibfnamefont {M.~W.}\ \bibnamefont {Doherty}}, \bibinfo {author} {\bibfnamefont {T.}~\bibnamefont {N{\"o}bauer}}, \bibinfo {author} {\bibfnamefont {F.}~\bibnamefont {Rempp}}, \bibinfo {author} {\bibfnamefont {G.}~\bibnamefont {Balasubramanian}}, \bibinfo {author} {\bibfnamefont {T.}~\bibnamefont {Wolf}}, \bibinfo {author} {\bibfnamefont {F.}~\bibnamefont {Reinhard}}, \bibinfo {author} {\bibfnamefont {L.~C.}\ \bibnamefont {Hollenberg}}, \bibinfo {author} {\bibfnamefont {F.}~\bibnamefont {Jelezko}}, \emph {et~al.},\ }\href@noop {} {\bibfield  {journal} {\bibinfo  {journal} {Nature Physics}\ }\textbf {\bibinfo {volume} {7}},\ \bibinfo {pages} {459} (\bibinfo {year} {2011})}\BibitemShut {NoStop}%
  \bibitem [{\citenamefont {Ajoy}\ and\ \citenamefont {Cappellaro}(2012)}]{Ajoy2012stable}%
    \BibitemOpen
    \bibfield  {author} {\bibinfo {author} {\bibfnamefont {A.}~\bibnamefont {Ajoy}}\ and\ \bibinfo {author} {\bibfnamefont {P.}~\bibnamefont {Cappellaro}},\ }\href@noop {} {\bibfield  {journal} {\bibinfo  {journal} {Physical Review A—Atomic, Molecular, and Optical Physics}\ }\textbf {\bibinfo {volume} {86}},\ \bibinfo {pages} {062104} (\bibinfo {year} {2012})}\BibitemShut {NoStop}%
  \bibitem [{\citenamefont {Ledbetter}\ \emph {et~al.}(2012)\citenamefont {Ledbetter}, \citenamefont {Jensen}, \citenamefont {Fischer}, \citenamefont {Jarmola},\ and\ \citenamefont {Budker}}]{ledbetter2012gyroscopes}%
    \BibitemOpen
    \bibfield  {author} {\bibinfo {author} {\bibfnamefont {M.}~\bibnamefont {Ledbetter}}, \bibinfo {author} {\bibfnamefont {K.}~\bibnamefont {Jensen}}, \bibinfo {author} {\bibfnamefont {R.}~\bibnamefont {Fischer}}, \bibinfo {author} {\bibfnamefont {A.}~\bibnamefont {Jarmola}},\ and\ \bibinfo {author} {\bibfnamefont {D.}~\bibnamefont {Budker}},\ }\href@noop {} {\bibfield  {journal} {\bibinfo  {journal} {Physical Review A—Atomic, Molecular, and Optical Physics}\ }\textbf {\bibinfo {volume} {86}},\ \bibinfo {pages} {052116} (\bibinfo {year} {2012})}\BibitemShut {NoStop}%
  \bibitem [{\citenamefont {Balasubramanian}\ \emph {et~al.}(2008)\citenamefont {Balasubramanian}, \citenamefont {Chan}, \citenamefont {Kolesov}, \citenamefont {Al-Hmoud}, \citenamefont {Tisler}, \citenamefont {Shin}, \citenamefont {Kim}, \citenamefont {Wojcik}, \citenamefont {Hemmer}, \citenamefont {Krueger} \emph {et~al.}}]{Balasubramanian08}%
    \BibitemOpen
    \bibfield  {author} {\bibinfo {author} {\bibfnamefont {G.}~\bibnamefont {Balasubramanian}}, \bibinfo {author} {\bibfnamefont {I.}~\bibnamefont {Chan}}, \bibinfo {author} {\bibfnamefont {R.}~\bibnamefont {Kolesov}}, \bibinfo {author} {\bibfnamefont {M.}~\bibnamefont {Al-Hmoud}}, \bibinfo {author} {\bibfnamefont {J.}~\bibnamefont {Tisler}}, \bibinfo {author} {\bibfnamefont {C.}~\bibnamefont {Shin}}, \bibinfo {author} {\bibfnamefont {C.}~\bibnamefont {Kim}}, \bibinfo {author} {\bibfnamefont {A.}~\bibnamefont {Wojcik}}, \bibinfo {author} {\bibfnamefont {P.~R.}\ \bibnamefont {Hemmer}}, \bibinfo {author} {\bibfnamefont {A.}~\bibnamefont {Krueger}}, \emph {et~al.},\ }\href@noop {} {\bibfield  {journal} {\bibinfo  {journal} {Nature}\ }\textbf {\bibinfo {volume} {455}},\ \bibinfo {pages} {648} (\bibinfo {year} {2008})}\BibitemShut {NoStop}%
  \bibitem [{\citenamefont {Maletinsky}\ \emph {et~al.}(2012)\citenamefont {Maletinsky}, \citenamefont {Hong}, \citenamefont {Grinolds}, \citenamefont {Hausmann}, \citenamefont {Lukin}, \citenamefont {Walsworth}, \citenamefont {Loncar},\ and\ \citenamefont {Yacoby}}]{Maletinsky12}%
    \BibitemOpen
    \bibfield  {author} {\bibinfo {author} {\bibfnamefont {P.}~\bibnamefont {Maletinsky}}, \bibinfo {author} {\bibfnamefont {S.}~\bibnamefont {Hong}}, \bibinfo {author} {\bibfnamefont {M.~S.}\ \bibnamefont {Grinolds}}, \bibinfo {author} {\bibfnamefont {B.}~\bibnamefont {Hausmann}}, \bibinfo {author} {\bibfnamefont {M.~D.}\ \bibnamefont {Lukin}}, \bibinfo {author} {\bibfnamefont {R.~L.}\ \bibnamefont {Walsworth}}, \bibinfo {author} {\bibfnamefont {M.}~\bibnamefont {Loncar}},\ and\ \bibinfo {author} {\bibfnamefont {A.}~\bibnamefont {Yacoby}},\ }\href@noop {} {\bibfield  {journal} {\bibinfo  {journal} {Nature nanotechnology}\ }\textbf {\bibinfo {volume} {7}},\ \bibinfo {pages} {320} (\bibinfo {year} {2012})}\BibitemShut {NoStop}%
  \bibitem [{\citenamefont {Kucsko}\ \emph {et~al.}(2013)\citenamefont {Kucsko}, \citenamefont {Maurer}, \citenamefont {Yao}, \citenamefont {Kubo}, \citenamefont {Noh}, \citenamefont {Lo}, \citenamefont {Park},\ and\ \citenamefont {Lukin}}]{Kucsko13}%
    \BibitemOpen
    \bibfield  {author} {\bibinfo {author} {\bibfnamefont {G.}~\bibnamefont {Kucsko}}, \bibinfo {author} {\bibfnamefont {P.~C.}\ \bibnamefont {Maurer}}, \bibinfo {author} {\bibfnamefont {N.~Y.}\ \bibnamefont {Yao}}, \bibinfo {author} {\bibfnamefont {M.}~\bibnamefont {Kubo}}, \bibinfo {author} {\bibfnamefont {H.~J.}\ \bibnamefont {Noh}}, \bibinfo {author} {\bibfnamefont {P.~K.}\ \bibnamefont {Lo}}, \bibinfo {author} {\bibfnamefont {H.}~\bibnamefont {Park}},\ and\ \bibinfo {author} {\bibfnamefont {M.~D.}\ \bibnamefont {Lukin}},\ }\href@noop {} {\bibfield  {journal} {\bibinfo  {journal} {Nature}\ }\textbf {\bibinfo {volume} {500}},\ \bibinfo {pages} {54} (\bibinfo {year} {2013})}\BibitemShut {NoStop}%
  \bibitem [{\citenamefont {Fujiwara}\ \emph {et~al.}(2020)\citenamefont {Fujiwara}, \citenamefont {Sun}, \citenamefont {Dohms}, \citenamefont {Nishimura}, \citenamefont {Suto}, \citenamefont {Takezawa}, \citenamefont {Oshimi}, \citenamefont {Zhao}, \citenamefont {Sadzak}, \citenamefont {Umehara} \emph {et~al.}}]{Fujiwara20}%
    \BibitemOpen
    \bibfield  {author} {\bibinfo {author} {\bibfnamefont {M.}~\bibnamefont {Fujiwara}}, \bibinfo {author} {\bibfnamefont {S.}~\bibnamefont {Sun}}, \bibinfo {author} {\bibfnamefont {A.}~\bibnamefont {Dohms}}, \bibinfo {author} {\bibfnamefont {Y.}~\bibnamefont {Nishimura}}, \bibinfo {author} {\bibfnamefont {K.}~\bibnamefont {Suto}}, \bibinfo {author} {\bibfnamefont {Y.}~\bibnamefont {Takezawa}}, \bibinfo {author} {\bibfnamefont {K.}~\bibnamefont {Oshimi}}, \bibinfo {author} {\bibfnamefont {L.}~\bibnamefont {Zhao}}, \bibinfo {author} {\bibfnamefont {N.}~\bibnamefont {Sadzak}}, \bibinfo {author} {\bibfnamefont {Y.}~\bibnamefont {Umehara}}, \emph {et~al.},\ }\href@noop {} {\bibfield  {journal} {\bibinfo  {journal} {Science advances}\ }\textbf {\bibinfo {volume} {6}},\ \bibinfo {pages} {eaba9636} (\bibinfo {year} {2020})}\BibitemShut {NoStop}%
  \bibitem [{\citenamefont {Fujiwara}\ and\ \citenamefont {Shikano}(2021)}]{Fujiwara21}%
    \BibitemOpen
    \bibfield  {author} {\bibinfo {author} {\bibfnamefont {M.}~\bibnamefont {Fujiwara}}\ and\ \bibinfo {author} {\bibfnamefont {Y.}~\bibnamefont {Shikano}},\ }\href@noop {} {\bibfield  {journal} {\bibinfo  {journal} {Nanotechnology}\ }\textbf {\bibinfo {volume} {32}},\ \bibinfo {pages} {482002} (\bibinfo {year} {2021})}\BibitemShut {NoStop}%
  \bibitem [{\citenamefont {Lesik}\ \emph {et~al.}(2019)\citenamefont {Lesik}, \citenamefont {Plisson}, \citenamefont {Toraille}, \citenamefont {Renaud}, \citenamefont {Occelli}, \citenamefont {Schmidt}, \citenamefont {Salord}, \citenamefont {Delobbe}, \citenamefont {Debuisschert}, \citenamefont {Rondin} \emph {et~al.}}]{Lesik19}%
    \BibitemOpen
    \bibfield  {author} {\bibinfo {author} {\bibfnamefont {M.}~\bibnamefont {Lesik}}, \bibinfo {author} {\bibfnamefont {T.}~\bibnamefont {Plisson}}, \bibinfo {author} {\bibfnamefont {L.}~\bibnamefont {Toraille}}, \bibinfo {author} {\bibfnamefont {J.}~\bibnamefont {Renaud}}, \bibinfo {author} {\bibfnamefont {F.}~\bibnamefont {Occelli}}, \bibinfo {author} {\bibfnamefont {M.}~\bibnamefont {Schmidt}}, \bibinfo {author} {\bibfnamefont {O.}~\bibnamefont {Salord}}, \bibinfo {author} {\bibfnamefont {A.}~\bibnamefont {Delobbe}}, \bibinfo {author} {\bibfnamefont {T.}~\bibnamefont {Debuisschert}}, \bibinfo {author} {\bibfnamefont {L.}~\bibnamefont {Rondin}}, \emph {et~al.},\ }\href@noop {} {\bibfield  {journal} {\bibinfo  {journal} {Science}\ }\textbf {\bibinfo {volume} {366}},\ \bibinfo {pages} {1359} (\bibinfo {year} {2019})}\BibitemShut {NoStop}%
  \bibitem [{\citenamefont {Hsieh}\ \emph {et~al.}(2019)\citenamefont {Hsieh}, \citenamefont {Bhattacharyya}, \citenamefont {Zu}, \citenamefont {Mittiga}, \citenamefont {Smart}, \citenamefont {Machado}, \citenamefont {Kobrin}, \citenamefont {H{\"o}hn}, \citenamefont {Rui}, \citenamefont {Kamrani} \emph {et~al.}}]{Hsieh19}%
    \BibitemOpen
    \bibfield  {author} {\bibinfo {author} {\bibfnamefont {S.}~\bibnamefont {Hsieh}}, \bibinfo {author} {\bibfnamefont {P.}~\bibnamefont {Bhattacharyya}}, \bibinfo {author} {\bibfnamefont {C.}~\bibnamefont {Zu}}, \bibinfo {author} {\bibfnamefont {T.}~\bibnamefont {Mittiga}}, \bibinfo {author} {\bibfnamefont {T.}~\bibnamefont {Smart}}, \bibinfo {author} {\bibfnamefont {F.}~\bibnamefont {Machado}}, \bibinfo {author} {\bibfnamefont {B.}~\bibnamefont {Kobrin}}, \bibinfo {author} {\bibfnamefont {T.}~\bibnamefont {H{\"o}hn}}, \bibinfo {author} {\bibfnamefont {N.}~\bibnamefont {Rui}}, \bibinfo {author} {\bibfnamefont {M.}~\bibnamefont {Kamrani}}, \emph {et~al.},\ }\href@noop {} {\bibfield  {journal} {\bibinfo  {journal} {Science}\ }\textbf {\bibinfo {volume} {366}},\ \bibinfo {pages} {1349} (\bibinfo {year} {2019})}\BibitemShut {NoStop}%
  \bibitem [{\citenamefont {Hilberer}\ \emph {et~al.}(2023)\citenamefont {Hilberer}, \citenamefont {Toraille}, \citenamefont {Dailledouze}, \citenamefont {Adam}, \citenamefont {Hanlon}, \citenamefont {Weck}, \citenamefont {Schmidt}, \citenamefont {Loubeyre},\ and\ \citenamefont {Roch}}]{Hilberer23}%
    \BibitemOpen
    \bibfield  {author} {\bibinfo {author} {\bibfnamefont {A.}~\bibnamefont {Hilberer}}, \bibinfo {author} {\bibfnamefont {L.}~\bibnamefont {Toraille}}, \bibinfo {author} {\bibfnamefont {C.}~\bibnamefont {Dailledouze}}, \bibinfo {author} {\bibfnamefont {M.-P.}\ \bibnamefont {Adam}}, \bibinfo {author} {\bibfnamefont {L.}~\bibnamefont {Hanlon}}, \bibinfo {author} {\bibfnamefont {G.}~\bibnamefont {Weck}}, \bibinfo {author} {\bibfnamefont {M.}~\bibnamefont {Schmidt}}, \bibinfo {author} {\bibfnamefont {P.}~\bibnamefont {Loubeyre}},\ and\ \bibinfo {author} {\bibfnamefont {J.-F.}\ \bibnamefont {Roch}},\ }\href@noop {} {\bibfield  {journal} {\bibinfo  {journal} {Physical Review B}\ }\textbf {\bibinfo {volume} {107}},\ \bibinfo {pages} {L220102} (\bibinfo {year} {2023})}\BibitemShut {NoStop}%
  \bibitem [{\citenamefont {Acosta}\ \emph {et~al.}(2010)\citenamefont {Acosta}, \citenamefont {Bauch}, \citenamefont {Ledbetter}, \citenamefont {Waxman}, \citenamefont {Bouchard},\ and\ \citenamefont {Budker}}]{Acosta10}%
    \BibitemOpen
    \bibfield  {author} {\bibinfo {author} {\bibfnamefont {V.~M.}\ \bibnamefont {Acosta}}, \bibinfo {author} {\bibfnamefont {E.}~\bibnamefont {Bauch}}, \bibinfo {author} {\bibfnamefont {M.~P.}\ \bibnamefont {Ledbetter}}, \bibinfo {author} {\bibfnamefont {A.}~\bibnamefont {Waxman}}, \bibinfo {author} {\bibfnamefont {L.-S.}\ \bibnamefont {Bouchard}},\ and\ \bibinfo {author} {\bibfnamefont {D.}~\bibnamefont {Budker}},\ }\href@noop {} {\bibfield  {journal} {\bibinfo  {journal} {Physical review letters}\ }\textbf {\bibinfo {volume} {104}},\ \bibinfo {pages} {070801} (\bibinfo {year} {2010})}\BibitemShut {NoStop}%
  \bibitem [{\citenamefont {Doherty}\ \emph {et~al.}(2014{\natexlab{b}})\citenamefont {Doherty}, \citenamefont {Struzhkin}, \citenamefont {Simpson}, \citenamefont {McGuinness}, \citenamefont {Meng}, \citenamefont {Stacey}, \citenamefont {Karle}, \citenamefont {Hemley}, \citenamefont {Manson}, \citenamefont {Hollenberg} \emph {et~al.}}]{Doherty14L}%
    \BibitemOpen
    \bibfield  {author} {\bibinfo {author} {\bibfnamefont {M.~W.}\ \bibnamefont {Doherty}}, \bibinfo {author} {\bibfnamefont {V.~V.}\ \bibnamefont {Struzhkin}}, \bibinfo {author} {\bibfnamefont {D.~A.}\ \bibnamefont {Simpson}}, \bibinfo {author} {\bibfnamefont {L.~P.}\ \bibnamefont {McGuinness}}, \bibinfo {author} {\bibfnamefont {Y.}~\bibnamefont {Meng}}, \bibinfo {author} {\bibfnamefont {A.}~\bibnamefont {Stacey}}, \bibinfo {author} {\bibfnamefont {T.~J.}\ \bibnamefont {Karle}}, \bibinfo {author} {\bibfnamefont {R.~J.}\ \bibnamefont {Hemley}}, \bibinfo {author} {\bibfnamefont {N.~B.}\ \bibnamefont {Manson}}, \bibinfo {author} {\bibfnamefont {L.~C.}\ \bibnamefont {Hollenberg}}, \emph {et~al.},\ }\href@noop {} {\bibfield  {journal} {\bibinfo  {journal} {Physical review letters}\ }\textbf {\bibinfo {volume} {112}},\ \bibinfo {pages} {047601} (\bibinfo {year} {2014}{\natexlab{b}})}\BibitemShut {NoStop}%
  \bibitem [{\citenamefont {Iv{\'a}dy}\ \emph {et~al.}(2014)\citenamefont {Iv{\'a}dy}, \citenamefont {Simon}, \citenamefont {Maze}, \citenamefont {Abrikosov},\ and\ \citenamefont {Gali}}]{Ivady14}%
    \BibitemOpen
    \bibfield  {author} {\bibinfo {author} {\bibfnamefont {V.}~\bibnamefont {Iv{\'a}dy}}, \bibinfo {author} {\bibfnamefont {T.}~\bibnamefont {Simon}}, \bibinfo {author} {\bibfnamefont {J.~R.}\ \bibnamefont {Maze}}, \bibinfo {author} {\bibfnamefont {I.}~\bibnamefont {Abrikosov}},\ and\ \bibinfo {author} {\bibfnamefont {A.}~\bibnamefont {Gali}},\ }\href@noop {} {\bibfield  {journal} {\bibinfo  {journal} {Physical Review B}\ }\textbf {\bibinfo {volume} {90}},\ \bibinfo {pages} {235205} (\bibinfo {year} {2014})}\BibitemShut {NoStop}%
  \bibitem [{\citenamefont {MacQuarrie}\ \emph {et~al.}(2013)\citenamefont {MacQuarrie}, \citenamefont {Gosavi}, \citenamefont {Jungwirth}, \citenamefont {Bhave},\ and\ \citenamefont {Fuchs}}]{Macquarrie13}%
    \BibitemOpen
    \bibfield  {author} {\bibinfo {author} {\bibfnamefont {E.}~\bibnamefont {MacQuarrie}}, \bibinfo {author} {\bibfnamefont {T.}~\bibnamefont {Gosavi}}, \bibinfo {author} {\bibfnamefont {N.}~\bibnamefont {Jungwirth}}, \bibinfo {author} {\bibfnamefont {S.}~\bibnamefont {Bhave}},\ and\ \bibinfo {author} {\bibfnamefont {G.}~\bibnamefont {Fuchs}},\ }\href@noop {} {\bibfield  {journal} {\bibinfo  {journal} {Physical review letters}\ }\textbf {\bibinfo {volume} {111}},\ \bibinfo {pages} {227602} (\bibinfo {year} {2013})}\BibitemShut {NoStop}%
  \bibitem [{\citenamefont {Meesala}\ \emph {et~al.}(2016)\citenamefont {Meesala}, \citenamefont {Sohn}, \citenamefont {Atikian}, \citenamefont {Kim}, \citenamefont {Burek}, \citenamefont {Choy},\ and\ \citenamefont {Lon{\v{c}}ar}}]{Meesala16}%
    \BibitemOpen
    \bibfield  {author} {\bibinfo {author} {\bibfnamefont {S.}~\bibnamefont {Meesala}}, \bibinfo {author} {\bibfnamefont {Y.-I.}\ \bibnamefont {Sohn}}, \bibinfo {author} {\bibfnamefont {H.~A.}\ \bibnamefont {Atikian}}, \bibinfo {author} {\bibfnamefont {S.}~\bibnamefont {Kim}}, \bibinfo {author} {\bibfnamefont {M.~J.}\ \bibnamefont {Burek}}, \bibinfo {author} {\bibfnamefont {J.~T.}\ \bibnamefont {Choy}},\ and\ \bibinfo {author} {\bibfnamefont {M.}~\bibnamefont {Lon{\v{c}}ar}},\ }\href@noop {} {\bibfield  {journal} {\bibinfo  {journal} {Physical Review Applied}\ }\textbf {\bibinfo {volume} {5}},\ \bibinfo {pages} {034010} (\bibinfo {year} {2016})}\BibitemShut {NoStop}%
  \bibitem [{\citenamefont {Bennett}\ \emph {et~al.}(2013)\citenamefont {Bennett}, \citenamefont {Yao}, \citenamefont {Otterbach}, \citenamefont {Zoller}, \citenamefont {Rabl},\ and\ \citenamefont {Lukin}}]{Bennett13}%
    \BibitemOpen
    \bibfield  {author} {\bibinfo {author} {\bibfnamefont {S.}~\bibnamefont {Bennett}}, \bibinfo {author} {\bibfnamefont {N.~Y.}\ \bibnamefont {Yao}}, \bibinfo {author} {\bibfnamefont {J.}~\bibnamefont {Otterbach}}, \bibinfo {author} {\bibfnamefont {P.}~\bibnamefont {Zoller}}, \bibinfo {author} {\bibfnamefont {P.}~\bibnamefont {Rabl}},\ and\ \bibinfo {author} {\bibfnamefont {M.~D.}\ \bibnamefont {Lukin}},\ }\href@noop {} {\bibfield  {journal} {\bibinfo  {journal} {Physical review letters}\ }\textbf {\bibinfo {volume} {110}},\ \bibinfo {pages} {156402} (\bibinfo {year} {2013})}\BibitemShut {NoStop}%
  \bibitem [{\citenamefont {Bayliss}\ \emph {et~al.}(2020)\citenamefont {Bayliss}, \citenamefont {Laorenza}, \citenamefont {Mintun}, \citenamefont {Kovos}, \citenamefont {Freedman},\ and\ \citenamefont {Awschalom}}]{Bayliss20}%
    \BibitemOpen
    \bibfield  {author} {\bibinfo {author} {\bibfnamefont {S.}~\bibnamefont {Bayliss}}, \bibinfo {author} {\bibfnamefont {D.}~\bibnamefont {Laorenza}}, \bibinfo {author} {\bibfnamefont {P.}~\bibnamefont {Mintun}}, \bibinfo {author} {\bibfnamefont {B.}~\bibnamefont {Kovos}}, \bibinfo {author} {\bibfnamefont {D.~E.}\ \bibnamefont {Freedman}},\ and\ \bibinfo {author} {\bibfnamefont {D.}~\bibnamefont {Awschalom}},\ }\href@noop {} {\bibfield  {journal} {\bibinfo  {journal} {Science}\ }\textbf {\bibinfo {volume} {370}},\ \bibinfo {pages} {1309} (\bibinfo {year} {2020})}\BibitemShut {NoStop}%
  \bibitem [{\citenamefont {Serrano}\ \emph {et~al.}(2022)\citenamefont {Serrano}, \citenamefont {Kuppusamy}, \citenamefont {Heinrich}, \citenamefont {Fuhr}, \citenamefont {Hunger}, \citenamefont {Ruben},\ and\ \citenamefont {Goldner}}]{Serrano22}%
    \BibitemOpen
    \bibfield  {author} {\bibinfo {author} {\bibfnamefont {D.}~\bibnamefont {Serrano}}, \bibinfo {author} {\bibfnamefont {S.~K.}\ \bibnamefont {Kuppusamy}}, \bibinfo {author} {\bibfnamefont {B.}~\bibnamefont {Heinrich}}, \bibinfo {author} {\bibfnamefont {O.}~\bibnamefont {Fuhr}}, \bibinfo {author} {\bibfnamefont {D.}~\bibnamefont {Hunger}}, \bibinfo {author} {\bibfnamefont {M.}~\bibnamefont {Ruben}},\ and\ \bibinfo {author} {\bibfnamefont {P.}~\bibnamefont {Goldner}},\ }\href@noop {} {\bibfield  {journal} {\bibinfo  {journal} {Nature}\ }\textbf {\bibinfo {volume} {603}},\ \bibinfo {pages} {241} (\bibinfo {year} {2022})}\BibitemShut {NoStop}%
  \bibitem [{\citenamefont {Harvey}\ and\ \citenamefont {Wasielewski}(2021)}]{Harvey21}%
    \BibitemOpen
    \bibfield  {author} {\bibinfo {author} {\bibfnamefont {S.~M.}\ \bibnamefont {Harvey}}\ and\ \bibinfo {author} {\bibfnamefont {M.~R.}\ \bibnamefont {Wasielewski}},\ }\href@noop {} {\bibfield  {journal} {\bibinfo  {journal} {Journal of the American Chemical Society}\ }\textbf {\bibinfo {volume} {143}},\ \bibinfo {pages} {15508} (\bibinfo {year} {2021})}\BibitemShut {NoStop}%
  \bibitem [{\citenamefont {Xie}\ \emph {et~al.}(2023)\citenamefont {Xie}, \citenamefont {Mao}, \citenamefont {Lin}, \citenamefont {Feng}, \citenamefont {Stoddart}, \citenamefont {Young},\ and\ \citenamefont {Wasielewski}}]{Xie23}%
    \BibitemOpen
    \bibfield  {author} {\bibinfo {author} {\bibfnamefont {F.}~\bibnamefont {Xie}}, \bibinfo {author} {\bibfnamefont {H.}~\bibnamefont {Mao}}, \bibinfo {author} {\bibfnamefont {C.}~\bibnamefont {Lin}}, \bibinfo {author} {\bibfnamefont {Y.}~\bibnamefont {Feng}}, \bibinfo {author} {\bibfnamefont {J.~F.}\ \bibnamefont {Stoddart}}, \bibinfo {author} {\bibfnamefont {R.~M.}\ \bibnamefont {Young}},\ and\ \bibinfo {author} {\bibfnamefont {M.~R.}\ \bibnamefont {Wasielewski}},\ }\href@noop {} {\bibfield  {journal} {\bibinfo  {journal} {Journal of the American Chemical Society}\ }\textbf {\bibinfo {volume} {145}},\ \bibinfo {pages} {14922} (\bibinfo {year} {2023})}\BibitemShut {NoStop}%
  \bibitem [{\citenamefont {Zadrozny}\ \emph {et~al.}(2017)\citenamefont {Zadrozny}, \citenamefont {Gallagher}, \citenamefont {Harris},\ and\ \citenamefont {Freedman}}]{Zadrozny17}%
    \BibitemOpen
    \bibfield  {author} {\bibinfo {author} {\bibfnamefont {J.~M.}\ \bibnamefont {Zadrozny}}, \bibinfo {author} {\bibfnamefont {A.~T.}\ \bibnamefont {Gallagher}}, \bibinfo {author} {\bibfnamefont {T.~D.}\ \bibnamefont {Harris}},\ and\ \bibinfo {author} {\bibfnamefont {D.~E.}\ \bibnamefont {Freedman}},\ }\href@noop {} {\bibfield  {journal} {\bibinfo  {journal} {Journal of the American Chemical Society}\ }\textbf {\bibinfo {volume} {139}},\ \bibinfo {pages} {7089} (\bibinfo {year} {2017})}\BibitemShut {NoStop}%
  \bibitem [{\citenamefont {Mena}\ \emph {et~al.}(2024)\citenamefont {Mena}, \citenamefont {Mann}, \citenamefont {Cowley-Semple}, \citenamefont {Bryan}, \citenamefont {Heutz}, \citenamefont {McCamey}, \citenamefont {Attwood},\ and\ \citenamefont {Bayliss}}]{Mena24}%
    \BibitemOpen
    \bibfield  {author} {\bibinfo {author} {\bibfnamefont {A.}~\bibnamefont {Mena}}, \bibinfo {author} {\bibfnamefont {S.~K.}\ \bibnamefont {Mann}}, \bibinfo {author} {\bibfnamefont {A.}~\bibnamefont {Cowley-Semple}}, \bibinfo {author} {\bibfnamefont {E.}~\bibnamefont {Bryan}}, \bibinfo {author} {\bibfnamefont {S.}~\bibnamefont {Heutz}}, \bibinfo {author} {\bibfnamefont {D.~R.}\ \bibnamefont {McCamey}}, \bibinfo {author} {\bibfnamefont {M.}~\bibnamefont {Attwood}},\ and\ \bibinfo {author} {\bibfnamefont {S.~L.}\ \bibnamefont {Bayliss}},\ }\href@noop {} {\bibfield  {journal} {\bibinfo  {journal} {arXiv preprint arXiv:2402.07572}\ } (\bibinfo {year} {2024})}\BibitemShut {NoStop}%
  \bibitem [{\citenamefont {Singh}\ \emph {et~al.}(2024{\natexlab{a}})\citenamefont {Singh}, \citenamefont {D'Souza}, \citenamefont {Zhong}, \citenamefont {Druga}, \citenamefont {Oshiro}, \citenamefont {Blankenship}, \citenamefont {Reimer}, \citenamefont {Breeze},\ and\ \citenamefont {Ajoy}}]{Singh24}%
    \BibitemOpen
    \bibfield  {author} {\bibinfo {author} {\bibfnamefont {H.}~\bibnamefont {Singh}}, \bibinfo {author} {\bibfnamefont {N.}~\bibnamefont {D'Souza}}, \bibinfo {author} {\bibfnamefont {K.}~\bibnamefont {Zhong}}, \bibinfo {author} {\bibfnamefont {E.}~\bibnamefont {Druga}}, \bibinfo {author} {\bibfnamefont {J.}~\bibnamefont {Oshiro}}, \bibinfo {author} {\bibfnamefont {B.}~\bibnamefont {Blankenship}}, \bibinfo {author} {\bibfnamefont {J.~A.}\ \bibnamefont {Reimer}}, \bibinfo {author} {\bibfnamefont {J.~D.}\ \bibnamefont {Breeze}},\ and\ \bibinfo {author} {\bibfnamefont {A.}~\bibnamefont {Ajoy}},\ }\href@noop {} {\bibfield  {journal} {\bibinfo  {journal} {arXiv preprint arXiv:2402.13898}\ } (\bibinfo {year} {2024}{\natexlab{a}})}\BibitemShut {NoStop}%
  \bibitem [{\citenamefont {Bridgman}(1964)}]{Bridgman64}%
    \BibitemOpen
    \bibfield  {author} {\bibinfo {author} {\bibfnamefont {P.~W.}\ \bibnamefont {Bridgman}},\ }in\ \href@noop {} {\emph {\bibinfo {booktitle} {Papers 32-58}}}\ (\bibinfo  {publisher} {Harvard University Press},\ \bibinfo {year} {1964})\ pp.\ \bibinfo {pages} {1851--1932}\BibitemShut {NoStop}%
  \bibitem [{\citenamefont {Jiang}\ and\ \citenamefont {Kloc}(2013)}]{Jiang13}%
    \BibitemOpen
    \bibfield  {author} {\bibinfo {author} {\bibfnamefont {H.}~\bibnamefont {Jiang}}\ and\ \bibinfo {author} {\bibfnamefont {C.}~\bibnamefont {Kloc}},\ }\href@noop {} {\bibfield  {journal} {\bibinfo  {journal} {MRS bulletin}\ }\textbf {\bibinfo {volume} {38}},\ \bibinfo {pages} {28} (\bibinfo {year} {2013})}\BibitemShut {NoStop}%
  \bibitem [{\citenamefont {Wu}\ \emph {et~al.}(2019)\citenamefont {Wu}, \citenamefont {Ng}, \citenamefont {Mirkhanov}, \citenamefont {Amirzhan}, \citenamefont {Nitnara},\ and\ \citenamefont {Oxborrow}}]{wu2019unraveling}%
    \BibitemOpen
    \bibfield  {author} {\bibinfo {author} {\bibfnamefont {H.}~\bibnamefont {Wu}}, \bibinfo {author} {\bibfnamefont {W.}~\bibnamefont {Ng}}, \bibinfo {author} {\bibfnamefont {S.}~\bibnamefont {Mirkhanov}}, \bibinfo {author} {\bibfnamefont {A.}~\bibnamefont {Amirzhan}}, \bibinfo {author} {\bibfnamefont {S.}~\bibnamefont {Nitnara}},\ and\ \bibinfo {author} {\bibfnamefont {M.}~\bibnamefont {Oxborrow}},\ }\href@noop {} {\bibfield  {journal} {\bibinfo  {journal} {The Journal of Physical Chemistry C}\ }\textbf {\bibinfo {volume} {123}},\ \bibinfo {pages} {24275} (\bibinfo {year} {2019})}\BibitemShut {NoStop}%
  \bibitem [{\citenamefont {Yang}\ \emph {et~al.}(2000)\citenamefont {Yang}, \citenamefont {Sloop}, \citenamefont {Weissman},\ and\ \citenamefont {Lin}}]{Yang00}%
    \BibitemOpen
    \bibfield  {author} {\bibinfo {author} {\bibfnamefont {T.-C.}\ \bibnamefont {Yang}}, \bibinfo {author} {\bibfnamefont {D.~J.}\ \bibnamefont {Sloop}}, \bibinfo {author} {\bibfnamefont {S.}~\bibnamefont {Weissman}},\ and\ \bibinfo {author} {\bibfnamefont {T.-S.}\ \bibnamefont {Lin}},\ }\href@noop {} {\bibfield  {journal} {\bibinfo  {journal} {The Journal of Chemical Physics}\ }\textbf {\bibinfo {volume} {113}},\ \bibinfo {pages} {11194} (\bibinfo {year} {2000})}\BibitemShut {NoStop}%
  \bibitem [{\citenamefont {Takeda}\ \emph {et~al.}(2002)\citenamefont {Takeda}, \citenamefont {Takegoshi},\ and\ \citenamefont {Terao}}]{Takeda02}%
    \BibitemOpen
    \bibfield  {author} {\bibinfo {author} {\bibfnamefont {K.}~\bibnamefont {Takeda}}, \bibinfo {author} {\bibfnamefont {K.}~\bibnamefont {Takegoshi}},\ and\ \bibinfo {author} {\bibfnamefont {T.}~\bibnamefont {Terao}},\ }\href@noop {} {\bibfield  {journal} {\bibinfo  {journal} {The Journal of chemical physics}\ }\textbf {\bibinfo {volume} {117}},\ \bibinfo {pages} {4940} (\bibinfo {year} {2002})}\BibitemShut {NoStop}%
  \bibitem [{\citenamefont {Singh}\ \emph {et~al.}(2024{\natexlab{b}})\citenamefont {Singh}, \citenamefont {D'Souza}, \citenamefont {Zhong}, \citenamefont {Druga}, \citenamefont {Oshiro}, \citenamefont {Blankenship}, \citenamefont {Reimer}, \citenamefont {Breeze},\ and\ \citenamefont {Ajoy}}]{singh2024room}%
    \BibitemOpen
    \bibfield  {author} {\bibinfo {author} {\bibfnamefont {H.}~\bibnamefont {Singh}}, \bibinfo {author} {\bibfnamefont {N.}~\bibnamefont {D'Souza}}, \bibinfo {author} {\bibfnamefont {K.}~\bibnamefont {Zhong}}, \bibinfo {author} {\bibfnamefont {E.}~\bibnamefont {Druga}}, \bibinfo {author} {\bibfnamefont {J.}~\bibnamefont {Oshiro}}, \bibinfo {author} {\bibfnamefont {B.}~\bibnamefont {Blankenship}}, \bibinfo {author} {\bibfnamefont {J.~A.}\ \bibnamefont {Reimer}}, \bibinfo {author} {\bibfnamefont {J.~D.}\ \bibnamefont {Breeze}},\ and\ \bibinfo {author} {\bibfnamefont {A.}~\bibnamefont {Ajoy}},\ }\href@noop {} {\bibfield  {journal} {\bibinfo  {journal} {arXiv preprint arXiv:2402.13898}\ } (\bibinfo {year} {2024}{\natexlab{b}})}\BibitemShut {NoStop}%
  \bibitem [{\citenamefont {Baer}\ and\ \citenamefont {Chronister}(1994{\natexlab{a}})}]{baer93temp}%
    \BibitemOpen
    \bibfield  {author} {\bibinfo {author} {\bibfnamefont {B.~J.}\ \bibnamefont {Baer}}\ and\ \bibinfo {author} {\bibfnamefont {E.~L.}\ \bibnamefont {Chronister}},\ }\href {https://doi.org/10.1063/1.466992} {\bibfield  {journal} {\bibinfo  {journal} {The Journal of Chemical Physics}\ }\textbf {\bibinfo {volume} {100}},\ \bibinfo {pages} {23} (\bibinfo {year} {1994}{\natexlab{a}})},\ \Eprint {https://arxiv.org/abs/https://pubs.aip.org/aip/jcp/article-pdf/100/1/23/19307282/23\_1\_online.pdf} {https://pubs.aip.org/aip/jcp/article-pdf/100/1/23/19307282/23\_1\_online.pdf} \BibitemShut {NoStop}%
  \bibitem [{\citenamefont {Rice}\ \emph {et~al.}(2013{\natexlab{a}})\citenamefont {Rice}, \citenamefont {Tham},\ and\ \citenamefont {Chronister}}]{Rice13}%
    \BibitemOpen
    \bibfield  {author} {\bibinfo {author} {\bibfnamefont {A.~P.}\ \bibnamefont {Rice}}, \bibinfo {author} {\bibfnamefont {F.~S.}\ \bibnamefont {Tham}},\ and\ \bibinfo {author} {\bibfnamefont {E.~L.}\ \bibnamefont {Chronister}},\ }\href@noop {} {\bibfield  {journal} {\bibinfo  {journal} {Journal of chemical crystallography}\ }\textbf {\bibinfo {volume} {43}},\ \bibinfo {pages} {14} (\bibinfo {year} {2013}{\natexlab{a}})}\BibitemShut {NoStop}%
  \bibitem [{\citenamefont {Croci}\ \emph {et~al.}(1993)\citenamefont {Croci}, \citenamefont {M{\"u}schenborn}, \citenamefont {G{\"u}ttler}, \citenamefont {Renn},\ and\ \citenamefont {Wild}}]{Croci93}%
    \BibitemOpen
    \bibfield  {author} {\bibinfo {author} {\bibfnamefont {M.}~\bibnamefont {Croci}}, \bibinfo {author} {\bibfnamefont {H.-J.}\ \bibnamefont {M{\"u}schenborn}}, \bibinfo {author} {\bibfnamefont {F.}~\bibnamefont {G{\"u}ttler}}, \bibinfo {author} {\bibfnamefont {A.}~\bibnamefont {Renn}},\ and\ \bibinfo {author} {\bibfnamefont {U.~P.}\ \bibnamefont {Wild}},\ }\href@noop {} {\bibfield  {journal} {\bibinfo  {journal} {Chemical physics letters}\ }\textbf {\bibinfo {volume} {212}},\ \bibinfo {pages} {71} (\bibinfo {year} {1993})}\BibitemShut {NoStop}%
  \bibitem [{\citenamefont {Jarmola}\ \emph {et~al.}(2012)\citenamefont {Jarmola}, \citenamefont {Acosta}, \citenamefont {Jensen}, \citenamefont {Chemerisov},\ and\ \citenamefont {Budker}}]{Jarmola12}%
    \BibitemOpen
    \bibfield  {author} {\bibinfo {author} {\bibfnamefont {A.}~\bibnamefont {Jarmola}}, \bibinfo {author} {\bibfnamefont {V.}~\bibnamefont {Acosta}}, \bibinfo {author} {\bibfnamefont {K.}~\bibnamefont {Jensen}}, \bibinfo {author} {\bibfnamefont {S.}~\bibnamefont {Chemerisov}},\ and\ \bibinfo {author} {\bibfnamefont {D.}~\bibnamefont {Budker}},\ }\href@noop {} {\bibfield  {journal} {\bibinfo  {journal} {Physical review letters}\ }\textbf {\bibinfo {volume} {108}},\ \bibinfo {pages} {197601} (\bibinfo {year} {2012})}\BibitemShut {NoStop}%
  \bibitem [{\citenamefont {Choe}\ \emph {et~al.}(2018)\citenamefont {Choe}, \citenamefont {Yoon}, \citenamefont {Lee}, \citenamefont {Oh}, \citenamefont {Lee}, \citenamefont {Kang}, \citenamefont {Lee},\ and\ \citenamefont {Lee}}]{CHOE20181066}%
    \BibitemOpen
    \bibfield  {author} {\bibinfo {author} {\bibfnamefont {S.}~\bibnamefont {Choe}}, \bibinfo {author} {\bibfnamefont {J.}~\bibnamefont {Yoon}}, \bibinfo {author} {\bibfnamefont {M.}~\bibnamefont {Lee}}, \bibinfo {author} {\bibfnamefont {J.}~\bibnamefont {Oh}}, \bibinfo {author} {\bibfnamefont {D.}~\bibnamefont {Lee}}, \bibinfo {author} {\bibfnamefont {H.}~\bibnamefont {Kang}}, \bibinfo {author} {\bibfnamefont {C.-H.}\ \bibnamefont {Lee}},\ and\ \bibinfo {author} {\bibfnamefont {D.}~\bibnamefont {Lee}},\ }\href {https://doi.org/https://doi.org/10.1016/j.cap.2018.06.002} {\bibfield  {journal} {\bibinfo  {journal} {Current Applied Physics}\ }\textbf {\bibinfo {volume} {18}},\ \bibinfo {pages} {1066} (\bibinfo {year} {2018})}\BibitemShut {NoStop}%
  \bibitem [{\citenamefont {Kraus}\ \emph {et~al.}(2014)\citenamefont {Kraus}, \citenamefont {Soltamov}, \citenamefont {Fuchs}, \citenamefont {Simin}, \citenamefont {Sperlich}, \citenamefont {Baranov}, \citenamefont {Astakhov},\ and\ \citenamefont {Dyakonov}}]{kraus2014magnetic}%
    \BibitemOpen
    \bibfield  {author} {\bibinfo {author} {\bibfnamefont {H.}~\bibnamefont {Kraus}}, \bibinfo {author} {\bibfnamefont {V.}~\bibnamefont {Soltamov}}, \bibinfo {author} {\bibfnamefont {F.}~\bibnamefont {Fuchs}}, \bibinfo {author} {\bibfnamefont {D.}~\bibnamefont {Simin}}, \bibinfo {author} {\bibfnamefont {A.}~\bibnamefont {Sperlich}}, \bibinfo {author} {\bibfnamefont {P.}~\bibnamefont {Baranov}}, \bibinfo {author} {\bibfnamefont {G.}~\bibnamefont {Astakhov}},\ and\ \bibinfo {author} {\bibfnamefont {V.}~\bibnamefont {Dyakonov}},\ }\href@noop {} {\bibfield  {journal} {\bibinfo  {journal} {Scientific reports}\ }\textbf {\bibinfo {volume} {4}},\ \bibinfo {pages} {5303} (\bibinfo {year} {2014})}\BibitemShut {NoStop}%
  \bibitem [{\citenamefont {Heimel}\ \emph {et~al.}(2003)\citenamefont {Heimel}, \citenamefont {Puschnig}, \citenamefont {Oehzelt}, \citenamefont {Hummer}, \citenamefont {Koppelhuber-Bitschnau}, \citenamefont {Porsch}, \citenamefont {Ambrosch-Draxl},\ and\ \citenamefont {Resel}}]{Heimel03}%
    \BibitemOpen
    \bibfield  {author} {\bibinfo {author} {\bibfnamefont {G.}~\bibnamefont {Heimel}}, \bibinfo {author} {\bibfnamefont {P.}~\bibnamefont {Puschnig}}, \bibinfo {author} {\bibfnamefont {M.}~\bibnamefont {Oehzelt}}, \bibinfo {author} {\bibfnamefont {K.}~\bibnamefont {Hummer}}, \bibinfo {author} {\bibfnamefont {B.}~\bibnamefont {Koppelhuber-Bitschnau}}, \bibinfo {author} {\bibfnamefont {F.}~\bibnamefont {Porsch}}, \bibinfo {author} {\bibfnamefont {C.}~\bibnamefont {Ambrosch-Draxl}},\ and\ \bibinfo {author} {\bibfnamefont {R.}~\bibnamefont {Resel}},\ }\href@noop {} {\bibfield  {journal} {\bibinfo  {journal} {Journal of Physics: Condensed Matter}\ }\textbf {\bibinfo {volume} {15}},\ \bibinfo {pages} {3375} (\bibinfo {year} {2003})}\BibitemShut {NoStop}%
  \bibitem [{\citenamefont {Baimova}(2024)}]{Baimova04}%
    \BibitemOpen
    \bibfield  {author} {\bibinfo {author} {\bibfnamefont {J.~A.}\ \bibnamefont {Baimova}},\ }\href@noop {} {\bibfield  {journal} {\bibinfo  {journal} {Nanomaterials}\ }\textbf {\bibinfo {volume} {14}} (\bibinfo {year} {2024})}\BibitemShut {NoStop}%
  \bibitem [{\citenamefont {Baer}\ and\ \citenamefont {Chronister}(1994{\natexlab{b}})}]{Baer94}%
    \BibitemOpen
    \bibfield  {author} {\bibinfo {author} {\bibfnamefont {B.~J.}\ \bibnamefont {Baer}}\ and\ \bibinfo {author} {\bibfnamefont {E.~L.}\ \bibnamefont {Chronister}},\ }\href {https://doi.org/10.1063/1.466992} {\bibfield  {journal} {\bibinfo  {journal} {The Journal of Chemical Physics}\ }\textbf {\bibinfo {volume} {100}},\ \bibinfo {pages} {23} (\bibinfo {year} {1994}{\natexlab{b}})},\ \Eprint {https://arxiv.org/abs/https://pubs.aip.org/aip/jcp/article-pdf/100/1/23/19307282/23\_1\_online.pdf} {https://pubs.aip.org/aip/jcp/article-pdf/100/1/23/19307282/23\_1\_online.pdf} \BibitemShut {NoStop}%
  \bibitem [{\citenamefont {Giannozzi}\ \emph {et~al.}(2009)\citenamefont {Giannozzi}, \citenamefont {Baroni}, \citenamefont {Bonini}, \citenamefont {Calandra}, \citenamefont {Car}, \citenamefont {Cavazzoni}, \citenamefont {Ceresoli}, \citenamefont {Chiarotti}, \citenamefont {Cococcioni}, \citenamefont {Dabo}, \citenamefont {Corso}, \citenamefont {Gironcoli}, \citenamefont {Fabris}, \citenamefont {Fratesi}, \citenamefont {Gebauer}, \citenamefont {Gerstmann}, \citenamefont {Gougoussis}, \citenamefont {Kokalj}, \citenamefont {Lazzeri}, \citenamefont {Martin-Samos}, \citenamefont {Marzari}, \citenamefont {Mauri}, \citenamefont {Mazzarello}, \citenamefont {Paolini}, \citenamefont {Pasquarello}, \citenamefont {Paulatto}, \citenamefont {Sbraccia}, \citenamefont {Scandolo}, \citenamefont {Sclauzero}, \citenamefont {Seitsonen}, \citenamefont {Smogunov}, \citenamefont {Umari},\ and\ \citenamefont {Wentzcovitch}}]{giannozzi_quantum_2009}%
    \BibitemOpen
    \bibfield  {author} {\bibinfo {author} {\bibfnamefont {P.}~\bibnamefont {Giannozzi}}, \bibinfo {author} {\bibfnamefont {S.}~\bibnamefont {Baroni}}, \bibinfo {author} {\bibfnamefont {N.}~\bibnamefont {Bonini}}, \bibinfo {author} {\bibfnamefont {M.}~\bibnamefont {Calandra}}, \bibinfo {author} {\bibfnamefont {R.}~\bibnamefont {Car}}, \bibinfo {author} {\bibfnamefont {C.}~\bibnamefont {Cavazzoni}}, \bibinfo {author} {\bibfnamefont {D.}~\bibnamefont {Ceresoli}}, \bibinfo {author} {\bibfnamefont {G.~L.}\ \bibnamefont {Chiarotti}}, \bibinfo {author} {\bibfnamefont {M.}~\bibnamefont {Cococcioni}}, \bibinfo {author} {\bibfnamefont {I.}~\bibnamefont {Dabo}}, \bibinfo {author} {\bibfnamefont {A.~D.}\ \bibnamefont {Corso}}, \bibinfo {author} {\bibfnamefont {S.~d.}\ \bibnamefont {Gironcoli}}, \bibinfo {author} {\bibfnamefont {S.}~\bibnamefont {Fabris}}, \bibinfo {author} {\bibfnamefont {G.}~\bibnamefont {Fratesi}}, \bibinfo {author} {\bibfnamefont {R.}~\bibnamefont {Gebauer}}, \bibinfo {author} {\bibfnamefont
    {U.}~\bibnamefont {Gerstmann}}, \bibinfo {author} {\bibfnamefont {C.}~\bibnamefont {Gougoussis}}, \bibinfo {author} {\bibfnamefont {A.}~\bibnamefont {Kokalj}}, \bibinfo {author} {\bibfnamefont {M.}~\bibnamefont {Lazzeri}}, \bibinfo {author} {\bibfnamefont {L.}~\bibnamefont {Martin-Samos}}, \bibinfo {author} {\bibfnamefont {N.}~\bibnamefont {Marzari}}, \bibinfo {author} {\bibfnamefont {F.}~\bibnamefont {Mauri}}, \bibinfo {author} {\bibfnamefont {R.}~\bibnamefont {Mazzarello}}, \bibinfo {author} {\bibfnamefont {S.}~\bibnamefont {Paolini}}, \bibinfo {author} {\bibfnamefont {A.}~\bibnamefont {Pasquarello}}, \bibinfo {author} {\bibfnamefont {L.}~\bibnamefont {Paulatto}}, \bibinfo {author} {\bibfnamefont {C.}~\bibnamefont {Sbraccia}}, \bibinfo {author} {\bibfnamefont {S.}~\bibnamefont {Scandolo}}, \bibinfo {author} {\bibfnamefont {G.}~\bibnamefont {Sclauzero}}, \bibinfo {author} {\bibfnamefont {A.~P.}\ \bibnamefont {Seitsonen}}, \bibinfo {author} {\bibfnamefont {A.}~\bibnamefont {Smogunov}}, \bibinfo {author}
    {\bibfnamefont {P.}~\bibnamefont {Umari}},\ and\ \bibinfo {author} {\bibfnamefont {R.~M.}\ \bibnamefont {Wentzcovitch}},\ }\href {https://doi.org/10.1088/0953-8984/21/39/395502} {\bibfield  {journal} {\bibinfo  {journal} {J. Phys.: Condens. Matter}\ }\textbf {\bibinfo {volume} {21}},\ \bibinfo {pages} {395502} (\bibinfo {year} {2009})}\BibitemShut {NoStop}%
  \bibitem [{\citenamefont {Rice}\ \emph {et~al.}(2013{\natexlab{b}})\citenamefont {Rice}, \citenamefont {Tham},\ and\ \citenamefont {Chronister}}]{rice_temperature_2013}%
    \BibitemOpen
    \bibfield  {author} {\bibinfo {author} {\bibfnamefont {A.~P.}\ \bibnamefont {Rice}}, \bibinfo {author} {\bibfnamefont {F.~S.}\ \bibnamefont {Tham}},\ and\ \bibinfo {author} {\bibfnamefont {E.~L.}\ \bibnamefont {Chronister}},\ }\href {https://doi.org/10.1007/s10870-012-0378-6} {\bibfield  {journal} {\bibinfo  {journal} {Journal of Chemical Crystallography}\ }\textbf {\bibinfo {volume} {43}},\ \bibinfo {pages} {14} (\bibinfo {year} {2013}{\natexlab{b}})}\BibitemShut {NoStop}%
  \bibitem [{\citenamefont {Rietveld}\ \emph {et~al.}(1970)\citenamefont {Rietveld}, \citenamefont {Maslen},\ and\ \citenamefont {Clews}}]{rietveld_x-ray_1970}%
    \BibitemOpen
    \bibfield  {author} {\bibinfo {author} {\bibfnamefont {H.~M.}\ \bibnamefont {Rietveld}}, \bibinfo {author} {\bibfnamefont {E.~N.}\ \bibnamefont {Maslen}},\ and\ \bibinfo {author} {\bibfnamefont {C.~J.~B.}\ \bibnamefont {Clews}},\ }\href {https://doi.org/10.1107/S0567740870003023} {\bibfield  {journal} {\bibinfo  {journal} {Acta Crystallographica Section B: Structural Crystallography and Crystal Chemistry}\ }\textbf {\bibinfo {volume} {26}},\ \bibinfo {pages} {693} (\bibinfo {year} {1970})},\ \bibinfo {note} {publisher: International Union of Crystallography}\BibitemShut {NoStop}%
  \bibitem [{\citenamefont {Güttler}\ \emph {et~al.}(1996)\citenamefont {Güttler}, \citenamefont {Croci}, \citenamefont {Renn},\ and\ \citenamefont {Wild}}]{guttler_single_1996}%
    \BibitemOpen
    \bibfield  {author} {\bibinfo {author} {\bibfnamefont {F.}~\bibnamefont {Güttler}}, \bibinfo {author} {\bibfnamefont {M.}~\bibnamefont {Croci}}, \bibinfo {author} {\bibfnamefont {A.}~\bibnamefont {Renn}},\ and\ \bibinfo {author} {\bibfnamefont {U.~P.}\ \bibnamefont {Wild}},\ }\href {https://doi.org/10.1016/0301-0104(96)00250-9} {\bibfield  {journal} {\bibinfo  {journal} {Chemical Physics}\ }\textbf {\bibinfo {volume} {211}},\ \bibinfo {pages} {421} (\bibinfo {year} {1996})}\BibitemShut {NoStop}%
  \bibitem [{\citenamefont {Hellman}\ \emph {et~al.}(2004)\citenamefont {Hellman}, \citenamefont {Razaznejad},\ and\ \citenamefont {Lundqvist}}]{hellman_potential-energy_2004}%
    \BibitemOpen
    \bibfield  {author} {\bibinfo {author} {\bibfnamefont {A.}~\bibnamefont {Hellman}}, \bibinfo {author} {\bibfnamefont {B.}~\bibnamefont {Razaznejad}},\ and\ \bibinfo {author} {\bibfnamefont {B.~I.}\ \bibnamefont {Lundqvist}},\ }\href {https://doi.org/10.1063/1.1645787} {\bibfield  {journal} {\bibinfo  {journal} {The Journal of Chemical Physics}\ }\textbf {\bibinfo {volume} {120}},\ \bibinfo {pages} {4593} (\bibinfo {year} {2004})}\BibitemShut {NoStop}%
  \bibitem [{\citenamefont {Harriman}(1978)}]{harriman1978}%
    \BibitemOpen
    \bibfield  {author} {\bibinfo {author} {\bibfnamefont {J.~E.}\ \bibnamefont {Harriman}},\ }in\ \href {https://doi.org/10.1016/B978-0-12-326350-6.50006-1} {\emph {\bibinfo {booktitle} {Theoretical {Foundations} of {Electron} {Spin} {Resonance}}}},\ \bibinfo {series} {Physical {Chemistry}: {A} {Series} of {Monographs}}, Vol.~\bibinfo {volume} {37},\ \bibinfo {editor} {edited by\ \bibinfo {editor} {\bibfnamefont {J.~E.}\ \bibnamefont {Harriman}}}\ (\bibinfo  {publisher} {Academic Press},\ \bibinfo {year} {1978})\ pp.\ \bibinfo {pages} {1--13}\BibitemShut {NoStop}%
  \bibitem [{\citenamefont {Reynhardt}\ \emph {et~al.}(1998)\citenamefont {Reynhardt}, \citenamefont {High},\ and\ \citenamefont {Van~Wyk}}]{Reynhardt98}%
    \BibitemOpen
    \bibfield  {author} {\bibinfo {author} {\bibfnamefont {E.}~\bibnamefont {Reynhardt}}, \bibinfo {author} {\bibfnamefont {G.}~\bibnamefont {High}},\ and\ \bibinfo {author} {\bibfnamefont {J.}~\bibnamefont {Van~Wyk}},\ }\href@noop {} {\bibfield  {journal} {\bibinfo  {journal} {The Journal of chemical physics}\ }\textbf {\bibinfo {volume} {109}},\ \bibinfo {pages} {8471} (\bibinfo {year} {1998})}\BibitemShut {NoStop}%
  \bibitem [{\citenamefont {Blankenship}\ \emph {et~al.}(2023)\citenamefont {Blankenship}, \citenamefont {Jones}, \citenamefont {Zhao}, \citenamefont {Singh}, \citenamefont {Sarkar}, \citenamefont {Li}, \citenamefont {Suh}, \citenamefont {Chen}, \citenamefont {Grigoropoulos},\ and\ \citenamefont {Ajoy}}]{Blankenship23}%
    \BibitemOpen
    \bibfield  {author} {\bibinfo {author} {\bibfnamefont {B.~W.}\ \bibnamefont {Blankenship}}, \bibinfo {author} {\bibfnamefont {Z.}~\bibnamefont {Jones}}, \bibinfo {author} {\bibfnamefont {N.}~\bibnamefont {Zhao}}, \bibinfo {author} {\bibfnamefont {H.}~\bibnamefont {Singh}}, \bibinfo {author} {\bibfnamefont {A.}~\bibnamefont {Sarkar}}, \bibinfo {author} {\bibfnamefont {R.}~\bibnamefont {Li}}, \bibinfo {author} {\bibfnamefont {E.}~\bibnamefont {Suh}}, \bibinfo {author} {\bibfnamefont {A.}~\bibnamefont {Chen}}, \bibinfo {author} {\bibfnamefont {C.~P.}\ \bibnamefont {Grigoropoulos}},\ and\ \bibinfo {author} {\bibfnamefont {A.}~\bibnamefont {Ajoy}},\ }\href@noop {} {\bibfield  {journal} {\bibinfo  {journal} {Nano Letters}\ }\textbf {\bibinfo {volume} {23}},\ \bibinfo {pages} {9272} (\bibinfo {year} {2023})}\BibitemShut {NoStop}%
  \bibitem [{\citenamefont {Blankenship}\ \emph {et~al.}(2024)\citenamefont {Blankenship}, \citenamefont {Li}, \citenamefont {Jones}, \citenamefont {Parashar}, \citenamefont {Zhao}, \citenamefont {Singh}, \citenamefont {Li}, \citenamefont {Arvin}, \citenamefont {Sarkar}, \citenamefont {Yang} \emph {et~al.}}]{Blankenship24}%
    \BibitemOpen
    \bibfield  {author} {\bibinfo {author} {\bibfnamefont {B.~W.}\ \bibnamefont {Blankenship}}, \bibinfo {author} {\bibfnamefont {J.}~\bibnamefont {Li}}, \bibinfo {author} {\bibfnamefont {Z.}~\bibnamefont {Jones}}, \bibinfo {author} {\bibfnamefont {M.}~\bibnamefont {Parashar}}, \bibinfo {author} {\bibfnamefont {N.}~\bibnamefont {Zhao}}, \bibinfo {author} {\bibfnamefont {H.}~\bibnamefont {Singh}}, \bibinfo {author} {\bibfnamefont {R.}~\bibnamefont {Li}}, \bibinfo {author} {\bibfnamefont {S.}~\bibnamefont {Arvin}}, \bibinfo {author} {\bibfnamefont {A.}~\bibnamefont {Sarkar}}, \bibinfo {author} {\bibfnamefont {R.}~\bibnamefont {Yang}}, \emph {et~al.},\ }\href@noop {} {\bibfield  {journal} {\bibinfo  {journal} {Nano Letters}\ }\textbf {\bibinfo {volume} {24}},\ \bibinfo {pages} {9711} (\bibinfo {year} {2024})}\BibitemShut {NoStop}%
  \bibitem [{\citenamefont {Ambrose}\ \emph {et~al.}(1991)\citenamefont {Ambrose}, \citenamefont {Basch{\'e}},\ and\ \citenamefont {Moerner}}]{ambrose1991detection}%
    \BibitemOpen
    \bibfield  {author} {\bibinfo {author} {\bibfnamefont {W.}~\bibnamefont {Ambrose}}, \bibinfo {author} {\bibfnamefont {T.}~\bibnamefont {Basch{\'e}}},\ and\ \bibinfo {author} {\bibfnamefont {W.}~\bibnamefont {Moerner}},\ }\href@noop {} {\bibfield  {journal} {\bibinfo  {journal} {The Journal of chemical physics}\ }\textbf {\bibinfo {volume} {95}},\ \bibinfo {pages} {7150} (\bibinfo {year} {1991})}\BibitemShut {NoStop}%
  \bibitem [{\citenamefont {Stannigel}\ \emph {et~al.}(2010)\citenamefont {Stannigel}, \citenamefont {Rabl}, \citenamefont {S{\o}rensen}, \citenamefont {Zoller},\ and\ \citenamefont {Lukin}}]{Stannigel10}%
    \BibitemOpen
    \bibfield  {author} {\bibinfo {author} {\bibfnamefont {K.}~\bibnamefont {Stannigel}}, \bibinfo {author} {\bibfnamefont {P.}~\bibnamefont {Rabl}}, \bibinfo {author} {\bibfnamefont {A.~S.}\ \bibnamefont {S{\o}rensen}}, \bibinfo {author} {\bibfnamefont {P.}~\bibnamefont {Zoller}},\ and\ \bibinfo {author} {\bibfnamefont {M.~D.}\ \bibnamefont {Lukin}},\ }\href@noop {} {\bibfield  {journal} {\bibinfo  {journal} {Physical review letters}\ }\textbf {\bibinfo {volume} {105}},\ \bibinfo {pages} {220501} (\bibinfo {year} {2010})}\BibitemShut {NoStop}%
  \bibitem [{\citenamefont {Guo}\ \emph {et~al.}(2023)\citenamefont {Guo}, \citenamefont {Stramma}, \citenamefont {Li}, \citenamefont {Roth}, \citenamefont {Huang}, \citenamefont {Jin}, \citenamefont {Parker}, \citenamefont {Arjona~Mart{\'\i}nez}, \citenamefont {Shofer}, \citenamefont {Michaels} \emph {et~al.}}]{Guo23}%
    \BibitemOpen
    \bibfield  {author} {\bibinfo {author} {\bibfnamefont {X.}~\bibnamefont {Guo}}, \bibinfo {author} {\bibfnamefont {A.~M.}\ \bibnamefont {Stramma}}, \bibinfo {author} {\bibfnamefont {Z.}~\bibnamefont {Li}}, \bibinfo {author} {\bibfnamefont {W.~G.}\ \bibnamefont {Roth}}, \bibinfo {author} {\bibfnamefont {B.}~\bibnamefont {Huang}}, \bibinfo {author} {\bibfnamefont {Y.}~\bibnamefont {Jin}}, \bibinfo {author} {\bibfnamefont {R.~A.}\ \bibnamefont {Parker}}, \bibinfo {author} {\bibfnamefont {J.}~\bibnamefont {Arjona~Mart{\'\i}nez}}, \bibinfo {author} {\bibfnamefont {N.}~\bibnamefont {Shofer}}, \bibinfo {author} {\bibfnamefont {C.~P.}\ \bibnamefont {Michaels}}, \emph {et~al.},\ }\href@noop {} {\bibfield  {journal} {\bibinfo  {journal} {Physical Review X}\ }\textbf {\bibinfo {volume} {13}},\ \bibinfo {pages} {041037} (\bibinfo {year} {2023})}\BibitemShut {NoStop}%
  \bibitem [{\citenamefont {Lee}\ \emph {et~al.}(2016)\citenamefont {Lee}, \citenamefont {Lee}, \citenamefont {Ovartchaiyapong}, \citenamefont {Minguzzi}, \citenamefont {Maze},\ and\ \citenamefont {Bleszynski~Jayich}}]{Lee16}%
    \BibitemOpen
    \bibfield  {author} {\bibinfo {author} {\bibfnamefont {K.~W.}\ \bibnamefont {Lee}}, \bibinfo {author} {\bibfnamefont {D.}~\bibnamefont {Lee}}, \bibinfo {author} {\bibfnamefont {P.}~\bibnamefont {Ovartchaiyapong}}, \bibinfo {author} {\bibfnamefont {J.}~\bibnamefont {Minguzzi}}, \bibinfo {author} {\bibfnamefont {J.~R.}\ \bibnamefont {Maze}},\ and\ \bibinfo {author} {\bibfnamefont {A.~C.}\ \bibnamefont {Bleszynski~Jayich}},\ }\href@noop {} {\bibfield  {journal} {\bibinfo  {journal} {Physical Review Applied}\ }\textbf {\bibinfo {volume} {6}},\ \bibinfo {pages} {034005} (\bibinfo {year} {2016})}\BibitemShut {NoStop}%
  \bibitem [{\citenamefont {Wasielewski}\ \emph {et~al.}(2020)\citenamefont {Wasielewski}, \citenamefont {Forbes}, \citenamefont {Frank}, \citenamefont {Kowalski}, \citenamefont {Scholes}, \citenamefont {Yuen-Zhou}, \citenamefont {Baldo}, \citenamefont {Freedman}, \citenamefont {Goldsmith}, \citenamefont {Goodson~III} \emph {et~al.}}]{Wasielewski20}%
    \BibitemOpen
    \bibfield  {author} {\bibinfo {author} {\bibfnamefont {M.~R.}\ \bibnamefont {Wasielewski}}, \bibinfo {author} {\bibfnamefont {M.~D.}\ \bibnamefont {Forbes}}, \bibinfo {author} {\bibfnamefont {N.~L.}\ \bibnamefont {Frank}}, \bibinfo {author} {\bibfnamefont {K.}~\bibnamefont {Kowalski}}, \bibinfo {author} {\bibfnamefont {G.~D.}\ \bibnamefont {Scholes}}, \bibinfo {author} {\bibfnamefont {J.}~\bibnamefont {Yuen-Zhou}}, \bibinfo {author} {\bibfnamefont {M.~A.}\ \bibnamefont {Baldo}}, \bibinfo {author} {\bibfnamefont {D.~E.}\ \bibnamefont {Freedman}}, \bibinfo {author} {\bibfnamefont {R.~H.}\ \bibnamefont {Goldsmith}}, \bibinfo {author} {\bibfnamefont {T.}~\bibnamefont {Goodson~III}}, \emph {et~al.},\ }\href@noop {} {\bibfield  {journal} {\bibinfo  {journal} {Nature Reviews Chemistry}\ }\textbf {\bibinfo {volume} {4}},\ \bibinfo {pages} {490} (\bibinfo {year} {2020})}\BibitemShut {NoStop}%
  \end{thebibliography}
\end{document}